\documentclass[a4paper,11pt]{article}
\usepackage{jcappub} 
\usepackage{lineno}
\usepackage{ulem}
\usepackage{float}
\usepackage{placeins} 

\usepackage{orcidlink}
\usepackage{xcolor}

\title{Large Field Polynomial Inflation in Palatini \boldmath {$f(R,\phi)$} Gravity}

\author[a, b]{Nilay Bostan \orcidlink{0000-0002-1129-4345},}
\author[c]{Canan Karahan \orcidlink{0000-0003-1218-0451},}
\author[d]{Ozan Sarg{\i}n \orcidlink{0000-0002-1014-3633}}

\affiliation[a]{Department of Physics and Astronomy, University of Iowa, 52242 Iowa City, IA, USA}
\affiliation[b]{Proton Accelerator Facility, Turkish Energy Nuclear and Mineral Research Agency, Nuclear Energy Research Institute, 06980, Ankara, Türkiye}
\affiliation[c]{National Defence University, Turkish Naval Academy, Department of Basic Sciences, 34942 Tuzla, {\.I}stanbul, Türkiye}
\affiliation[d]{Sabanc{\i} University, Faculty of Engineering and Natural Sciences, 34956 Tuzla, {\.I}stanbul, Türkiye}
\emailAdd{nilay.bostan@tenmak.gov.tr}
\emailAdd{ckarahan@dho.edu.tr}
\emailAdd{ozan.sargin@sabanciuniv.edu}

\abstract{In this paper, we employ the Palatini formalism to investigate the dynamics of large-field inflation using a renormalizable polynomial inflaton potential in the context of $f(R,\phi)$ gravity. Assuming instant reheating, we make a comparative analysis of large-field polynomial inflation (PI). We first consider the minimal and non-minimal coupling of inflaton in $R$ gravity, and then we continue with the minimally and non-minimally coupled inflaton in $f(R,\phi)$ gravity. We scan the parameter space for the inflationary predictions ($n_s$ and $r$) consistent with the Planck and BICEP/Keck 2018 results as well as the sensitivity forecast of the future CMB-S4 and depict the compliant regions in the $\phi_0-\beta$ plane where $\phi_0$ and $\beta$ are two parameters of polynomial inflation model which control the saddle point of the potential and the flatness in the vicinity of this point respectively. We find that a substantial portion of the parameter space aligns with the observational data. 
}

\keywords{inflation, inflection point inflation, Palatini formalism, modified gravity}

\begin{document}
\maketitle
\flushbottom
\section{Introduction} \label{sec:intro}

Cosmic inflation posits that the universe underwent an extremely rapid exponential 
expansion in its very early stages, right after the Big Bang, during the first tiny fraction of a second \cite{Starobinsky:1980te, Guth:1980zm, Linde:1981mu, Albrecht:1982wi}. This period of inflation helps explain the uniformity of the Cosmic Microwave Background (CMB) radiation and the universe's large-scale structure \cite{Mukhanov:1981xt, guth, hawking}.

 There is a significant body of literature on inflationary cosmology \cite{Ellis:2023, Martin:2013tda, Odintsov:2023weg}. Amongst the plethora of inflationary models, the single-field models are the simplest ones, where a single scalar field slowly rolls down a sufficiently flat potential profile \cite{Lyth:2009zz}. Those profiles are generically monomials of the inflaton with a power of two or four, so they are bounded from below and renormalizable. The problem with these monomial potentials is they are sufficiently flat only at fairly large field values. This results in the overproduction of tensor modes and consequently in a large tensor-to-scalar ratio $r$ which measures the strength of the tensor perturbations (gravitational waves) relative to the scalar perturbations (density fluctuations). Thus, those models are ruled out after the release of Planck and  BICEP/Keck 2018 results \cite{Aghanim:2018eyx, BICEP:2021xfz}.

The next simplest model involves a single real scalar field again. However, it has a renormalizable potential: a fourth-degree polynomial instead of a monomial. This model is dubbed the polynomial inflation (PI) \cite{Kobayashi:2014, Drees_2024, Destri:2008, Wolfson:2019, Bernal:2021}. It is similar to the inflection point inflation scenario \cite{Enqvist:2010vd, Hotchkiss:2011am, Dimopoulos:2017xox, Okada:2016ssd, Okada:2017cvy, Okada:2019yne, Bai:2020zil}. In standard inflection point inflation, the inflaton potential has an inflection point, a point where the second derivative of the potential changes sign, but the first derivative remains small, and near this point, the potential can be approximated by a cubic function. Similarly, in PI, the polynomial potential has a saddle point, where both the first and second derivatives of the potential vanish. In practice, the location of this (would-be) saddle point is used as a free parameter, and a second free parameter controls the slope of the potential profile to modulate the flatness of the inflaton excursion zone.    

PI has been the subject of many analyses since the '90s because it is possible to encounter this type of potential in both minimal supersymmetric standard model (MSSM) and string theory \cite{Hodges:1989dw, Destri:2007pv, Aslanyan:2015hmi, Allahverdi:2006iq, Nakayama:2013jka, Nakayama:2013txa, Kallosh:2014xwa, Li:2014zfa, Gao:2015yha,Musoke:2017frr, Linde:2007jn}. PI can bear fruit in string theory in the sense that, in the limit where inflation occurs at field values that are very far away from the saddle point the potential approximates a purely quartic one. This in turn leads to eternal inflation which helps justify the weak anthropic principle and relax the initial conditions problem of inflation itself \cite{linde_initial, brandenberger, garfinkle}.

PI has been analyzed for both small \cite{Drees:2021wgd} and large field values \cite{Drees:2022aea, Zhang:2024ldx} considering the latest measurements and utilizing the metric formalism. Unlike the analyses conducted in these papers, we work out polynomial inflation in Palatini formalism in different cosmological settings like minimal and non-minimal inflaton coupling to standard metric-Palatini $R$ gravity and metric-Palatini $f(R,\phi)$ gravity.

The paper is structured as follows. Section \ref{Sec.2} provides a brief overview of the difference between metric and Palatini formulation of $f(R,\phi)$ concerning inflation. Section \ref{Sec.3} discusses the fundamentals of polynomial inflation in four different cosmological settings: minimal and non-minimal coupling of inflaton in standard metric-Palatini $R$ gravity (standard Einstein theory of gravity) or $f(R,\phi)$ metric-Palatini gravity. Section \ref{Sec.4} analyzes the inflationary dynamics and compares the model's predictions with observational data. Section \ref{conc} concludes with a summary and discussion of the implications and potential future research directions.

\section{\boldmath{$f(R,\phi)$} gravity} \label{Sec.2}

After the initial proposal of general relativity (GR), various attempts were undertaken to devise alternative theories of gravity. Some of the most captivating endeavors sought to formulate a more encompassing theory by relaxing one or more of the assumptions of GR. These included challenging the assumption that the only degrees of freedom of the gravitational field are those of the metric and the assumption that the gravitational lagrangian should be a linear function of the scalar curvature.

An extension of general relativity that garnered early interest involved modifying the gravitational action to allow for looser dependence on curvature invariants. One specific category of such theories, where the lagrangian is a general function of the scalar curvature, is known as $f(R)$ gravity \cite{buh,De Felice2010,Sotiriou2010,Capozziello2011,Clifton2012,Nojiri2017a,Nojiri:2010wj}. 

$f(R)$ theories are intrinsically non-minimal forms of gravity. They are non-minimal in the sense that they can either be described by higher order terms in Ricci scalar or by a scalar field non-minimally coupled to gravity. Either way, there is more than one choice for the dynamical degrees of freedom. 

One such choice is the metric formulation where the metric tensor is taken to be the only dynamical degree of freedom and the connection is fixed to be the Levi-Civita connection. In metric formulation, after applying the scalar-tensor theory and $f(R)$ gravity equivalence, a dynamical scalar field $\chi$ arises naturally via the conformal factor in the Einstein frame. Therefore, $f(R)$ theories in metric formulation could be suitable for inflation since they contain this scalar degree of freedom (called scalaron) that can act as the inflaton \cite{Bostan:2023hjp, Myrzakulov:2015qaa}. The Starobinsky model is a prime example of such models \cite{Starobinsky:1980te}. However, if the theory involves another dynamical scalar field $\phi$ (in such a case it is called an $f(R,\phi)$ gravity), both scalars $\phi$ and $\chi$ can play the role of the inflaton and one faces a two field inflation scenario \cite{Mori:2017caa}. 

The second choice for the set of degrees of freedom is known as the Palatini formalism of $f(R)$ gravity. By altering the variational principle and treating the metric and connection as independent quantities, a different theory can be derived from the same action. This means that the connection is no longer required to be the Levi-Civita connection of the metric, and the resulting theory may not be a metric one  \cite{ferraris,papapetrou,kunstatter,ferraris2,sotiriou}. The advantage of the Palatini over the metric formulation is that one still obtains a single-field inflationary setup even though there may exist two scalar fields $\phi$ and $\chi$ after the theory is imported to the Einstein frame. This is because the auxiliary field $\chi$ becomes non-dynamical in the Palatini formalism and its equation of motion constitutes just a constraint over the system. The literature includes numerous studies on inflationary models in Palatini $f(R,\phi)$ gravity see  \cite{Antoniadis:2018ywb, Dioguardi:2021fmr, Lahanas:2022mng, Dioguardi:2022oqu, AlHallak:2022gbv, Gialamas:2023flv, Bostan:2024ugi, Gialamas:2019nly, Gialamas_2020, Gialamas:2020vto, Gialamas:2021enw, Allemandi:2005qs, Bauer:2008zj, Bostan:2020pnb, Bostan:2019uvv, Tenkanen:2017jih, Rasanen:2017ivk, Rubio:2019ypq, Tenkanen:2019jiq, Dimopoulos:2022rdp, Antoniadis:2018yfq, Karam:2021sno, Karam:2018mft} and references therein.

In the subsequent sections, we will investigate polynomial inflation in the $f(R,\phi)$ gravity setting using Palatini formulation. But first, let us proceed with an overview of it in the next section.

\section{Polynomial inflation} \label{Sec.3}

In PI, a polynomial function of the inflaton describes the potential energy driving the inflationary expansion \cite{Bernal:2024ykj}. The generic form of the renormalizable potential \footnote{Higher-order terms could be added to the effective potential. However, due to insufficient information about the boundaries of the effective description and the analysis conducted in \cite{Jinno_2020},  which indicates that it is necessary to suppress these terms in the context of Higgs inflation, we have opted to maintain the potential at the renormalizable level.} is

\begin{align}	\label{inflaton_potential}
    V(\phi) = b\, \phi^2 + c\, \phi^3 + d\, \phi^4 \,.
\end{align}
Here, $\phi$ is the inflaton field; $b$, $c$ and $d$ are some constants. These constants have to satisfy certain conditions. The constant $d$ multiplies the term with the highest power. It has to be positive so that $V(\phi)$ is bounded from below. The constant in front of the quadratic term must be positive such that the potential (\ref{inflaton_potential}) develops a minimum at the origin. Additionally, the constant $c$ of the cubic term is chosen to be negative to restrict inflation to the positive field space. The linear term is absent in $V(\phi)$ because it can be absorbed into the other terms by a simple field redefinition. The prime advantage of (\ref{inflaton_potential}) over the monomial models is that through fine-tuning the composition of the terms in it, one obtains a fairly flat region close to the saddle point. The location of the saddle point is determined by setting the first and second derivatives 

\begin{align}
V^{\prime}(\phi) =  2 b\, \phi + 3 c\, \phi^2 + 4d\, \phi^3\,; \ \ \
V^{\prime \prime}(\phi) =  2 b  + 6 c\, \phi + 12 d\, \phi^2\,
\end{align}
to zero. This requires 
\begin{align} \label{correlation}
\phi_0 = - \frac{3c}{8d}, \qquad  b = \frac{9c^2}{32d}\,,
\end{align}
and here we see that the ratio $c/d$ determines the location of the saddle point. Having determined the saddle point $\phi_0$; it is advantageous to parametrize (\ref{inflaton_potential}) in terms of $\phi_0$, $d$ and a new parameter $\beta$ which controls the slope near $\phi_0$. The reparametrized form of the potential (\ref{inflaton_potential}) reads

\begin{eqnarray} \label{pot_rep}
V(\phi)=d\left[\phi^4- \frac{8\phi_0}{3}(1-\beta)\phi^3+2\:\phi_0^2 \: \phi^2\right] \; .
\end{eqnarray}
The deviation factor $(1-\beta)$ has been introduced to the reparametrized form (\ref{pot_rep}) so that the flatness of the potential profile can be modulated as needed for successful inflation. The overall multiplicative factor $d$ affects only the normalization of density perturbations. On the other hand, the parameter $\beta$ is consequential and should be non-negative to ensure a hot Big Bang history for the universe. Because, if $\beta< 0$ the inflaton would have a false vacuum at $\phi > \phi_0$ where it could get stuck.

\begin{figure}[tbp]
\centering
    \includegraphics[width=0.63\textwidth]{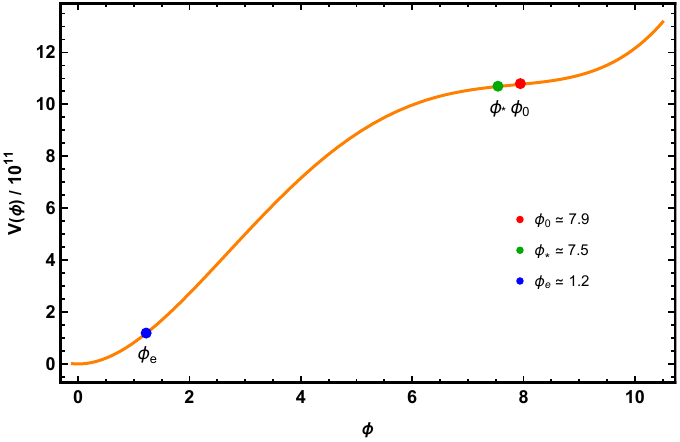} 
    \caption{\label{fig:potform} Generic plot of PI potential for $\beta=6.6\times10^{-3}$. The red dot corresponds to the saddle point ($\phi_0$). The green ($\phi_* $) and blue ($\phi_e $) dots designate the field value when the pivot scale exits the horizon and the end of inflation respectively.}
\end{figure}

In this work, we consider a large field inflation scenario where the saddle point is taken to be $\phi_0 \geq 1$ such that inflation is confined to trans-Planckian field values. Figure \ref{fig:potform} illustrates this with a representative plot for the PI potential profile. In this figure, the red dot corresponds to the saddle point $\phi_0$. The green and blue dots designate the inflaton field value at the moment the cosmological pivot scale ($k_*$) first exits the horizon ($\phi_* $) and at the end of inflation ($\phi_e $) respectively. It is evident that the inflationary excursion takes place in the trans-Planckian domain.

In the rest of this section, we will conduct a step-by-step analysis of polynomial inflation to reveal the effects of various terms and couplings in metric-Palatini $f(R,\phi)$ gravity. The steps that constitute this venture are

{\centering 
\begin{itemize}
    \item Sec. \ref{subsubsec. 3.1.1}: \textit{``Minimal coupling in $R$ gravity"} (which corresponds to the scenario that inflaton is coupled to linear gravity (gravity linear in the curvature) via only metric)     
        \item Sec. \ref{subsubsec. 3.1.2}: \textit{``Non-minimal coupling in $R$ gravity"} (which corresponds to the scenario where inflaton is coupled to gravity via the lagrangian term $\xi \phi^2 R$) 
    \item Sec. \ref{subsubsec. 3.2.1}: \textit{``Minimal coupling in $f(R,\phi)$ gravity"} ($R+R^2$ form of $f(R)$ theory and the inflaton is coupled to gravity via only metric)
     \item Sec. \ref{subsubsec. 3.2.2}: \textit{``Non-minimal coupling in $f(R,\phi)$  gravity"} ($R+R^2$ form of an $f(R)$ gravity to which the inflaton couples via $\xi \phi^2 R$).
   \end{itemize}
}

\noindent {The details of the various cases mentioned will be discussed next.}


\subsection{Polynomial inflation in metric-Palatini \boldmath{$R$} gravity} \label{subsec. 3.1}

\subsubsection{Minimal coupling} \label{subsubsec. 3.1.1}

This is the simplest case that we consider. The inflationary action is just comprised of the Einstein-Hilbert term plus the kinetic and potential terms of the inflaton. The action reads

\begin{eqnarray}\label{minimal-linear-case}
S_{ML}=\int d^4x \sqrt{-g} \Bigg\{
\frac{1}{2}M_{Pl}^2 R -\frac{1}{2} \partial_\mu\phi \partial^\mu \phi  -V(\phi)\Bigg\},
\end{eqnarray}
where $V(\phi)$ is the reparametrized potential given in (\ref{pot_rep}). 

Even though it is based on the Palatini formalism, this case corresponds to the standard Einstein theory of gravity at the equation of motion level. As seen in Sec. \ref{subsubsec.4.1.1}, the inflationary predictions are very close to the ones in \cite{Drees:2022aea}.


\subsubsection{Non-minimal coupling} \label{subsubsec. 3.1.2}

The non-minimal coupling involves the inflaton interacting with gravity through both the metric and the explicit ($\phi^2 R$) Ricci scalar interaction term. In Jordan frame, the action takes the form
\begin{eqnarray}\label{nonminimal-case}
S_{NL}=\int d^4x \sqrt{-g} \Bigg\{
\frac{1}{2}M_{Pl}^2 R + \frac{1}{2}\xi\phi^2 R -\frac{1}{2} \partial_\mu\phi \partial^\mu \phi  -V(\phi)\Bigg\}.
\end{eqnarray}
By performing the conformal transformation $g_{\mu \nu}\rightarrow Z_1(\phi)g_{\mu\nu}$ where 
\begin{eqnarray}
Z_1(\phi)=\left(1+\frac{\xi}{M_{Pl}^2} \phi^2\right)
\end{eqnarray}
the action takes the Einstein frame form 
\begin{eqnarray}\label{nonminimal-case_EF}
S_{NL}=\int d^4x \sqrt{-g} \Bigg\{
\frac{1}{2}M_{Pl}^2 R -\frac{1}{2} \frac{\partial_\mu\phi \partial^\mu \phi}{Z_1(\phi)}  -\frac{V(\phi)}{Z_1(\phi)^2}\Bigg\},
\end{eqnarray}
and the potential in Einstein frame reads
\begin{eqnarray}\label{nonminimal-case_pot_EF}
V_{NL}(\phi)= \frac{V(\phi)}{Z_1(\phi)^2},
\end{eqnarray}
where $V(\phi)$ is the reparametrized potential given in (\ref{pot_rep}).


\subsection{Polynomial inflation in metric-Palatini \boldmath{$f(R,\phi)$} gravity} \label{subsec. 3.2}
\subsubsection{Minimal coupling} \label{subsubsec. 3.2.1}

In this section, we couple the Starobinsky action to matter in a minimal manner and without introducing a new mass scale. Therefore, we introduce the scalar $\phi$ with a classically scale invariant minimal coupling to gravity through the action
\begin{eqnarray}\label{minimal-quad-case}
S_{MQ}=\int d^4x \sqrt{-g} \Bigg\{
\frac{1}{2}M_{Pl}^2 R - \frac{1}{2} \alpha R^2 -\frac{1}{2} \partial_\mu\phi \partial^\mu \phi  -V(\phi)\Bigg\}.
\end{eqnarray}
Written in terms of the auxiliary field $\chi$ the action (\ref{minimal-quad-case}) becomes
\begin{eqnarray}\label{minimal-quad-case_nm}
S_{MQ} \rightarrow \int d^4x \sqrt{-g} \Bigg\{
\frac{1}{2}M_{Pl}^2 F(\chi) R - \frac{1}{2} \partial_\mu\phi \partial^\mu \phi  
-\left(V(\phi)-\frac{\alpha}{8}\chi^4\right)\Bigg\}
\end{eqnarray}
where  
\begin{eqnarray}\label{conformal-factor_mq}
F(\chi)=\left(1 -\frac{\alpha}{M_{Pl}^2}\chi^2\right).
\end{eqnarray}
Next, we Weyl rescale the metric $g_{\mu \nu}\rightarrow F(\chi)g_{\mu \nu}$ so that (\ref{minimal-quad-case_nm}) can be cast into the Einstein frame form
\begin{eqnarray}\label{minimal-quad-case_m} 
S_{MQ}=\int d^4x \sqrt{-g} \Bigg\{
\frac{1}{2}M_{Pl}^2 R -\frac{1}{2} \left(\frac{\nabla \phi}{\sqrt{F(\chi)}}\right)^2  -\frac{V_J(\phi,\chi)}{F(\chi)^2}\Bigg\},
\end{eqnarray}
where $V_J(\phi,\chi)=V(\phi)-\alpha \chi^4/8$.

Variation of (\ref{minimal-quad-case_m}) with respect to $\chi$ allow for solving the auxiliary field $\chi$ in terms of the inflaton $\phi$

\begin{eqnarray}\label{minimal-quad-auxiliary}
\chi^2=\frac{4V(\phi)+(\nabla \phi)^2}{\frac{1}{2}M_{Pl}^2 + \frac{\alpha}{M_{Pl}^2}(\nabla \phi)^2}.
\end{eqnarray}
Applying slow-roll approximation $(\nabla \phi)^2 \ll \phi^2 $ to (\ref{minimal-quad-auxiliary}) we get
\begin{eqnarray}\label{minimal-quad-auxiliary_appr}
\chi^2=\frac{8V(\phi)}{M_{Pl}^2},
\end{eqnarray}
Substituting (\ref{minimal-quad-auxiliary_appr}) into (\ref{minimal-quad-case_m}) takes us to, 

\begin{eqnarray}\label{minimal-quad-case_m_EF}
S_{MQ}=\int d^4x \sqrt{-g} \Bigg\{
\frac{1}{2}M_{Pl}^2 R -\frac{1}{2} \left(\frac{\nabla \phi}{\sqrt{Z_2(\phi)}}\right)^2  -\frac{V_J(\phi)}{Z_2(\phi)^2}\Bigg\},
\end{eqnarray}
where 
\begin{eqnarray} \label{VJ_MQ_EF}
V_J(\phi)=V(\phi)- 8 \alpha\left[\frac{V(\phi)}{M_{Pl}^2}\right]^2,
\end{eqnarray}
and 
\begin{eqnarray} \label{f__MQ_EF}
Z_2(\phi)=1-\frac{8\alpha V(\phi)}{M_{Pl}^4 }.
\end{eqnarray}
Then, the resulting Einstein frame potential $V_{MQ}(\phi)=V_J(\phi)/Z_{2}(\phi)^2$ is

\begin{eqnarray} \label{V_MQ_EF}
V_{MQ}(\phi)=\frac{V(\phi)-\frac{8\alpha V(\phi)^2}{M_{Pl}^4 }}{Z_{2}(\phi)^2}\;,
\end{eqnarray}
which simplifies to 
\begin{eqnarray} \label{V_MQ_EF_simp}
V_{MQ}(\phi)=\frac{V(\phi)}{Z_{2}(\phi)}\;,
\end{eqnarray}
where $V(\phi)$ is the reparametrized potential given in (\ref{pot_rep}).

Comparing (\ref{V_MQ_EF_simp}) with $V(\phi)$ (which is the inflaton potential in Sec. \ref{subsubsec. 3.1.1}) we notice that, even though the inflaton is minimally coupled in both cases, the presence of the quadratic Ricci scalar term alters the inflaton potential drastically. We will witness the ramifications of this difference in Sec. \ref{subsubsec.4.2.1}.


\subsubsection{Non-minimal coupling} \label{subsubsec. 3.2.2}

The inflationary dynamics in this case are controlled by the action 

\begin{eqnarray}\label{symmergent-case}
S_{NQ}=\int d^4x \sqrt{-g} \Bigg\{
\frac{1}{2}M_{Pl}^2 R + \frac{1}{2}\xi\phi^2 R - \frac{1}{2} \alpha R^2 -\frac{1}{2} \partial_\mu\phi \partial^\mu \phi  -V(\phi)\Bigg\}.
\end{eqnarray}
This gravitational action is cast into the scalar-tensor equivalent form. Then, 
 it becomes 
\begin{eqnarray}\label{symmergent-case_nm}
S_{NQ} \rightarrow \int d^4x \sqrt{-g} \Bigg\{
\frac{1}{2}M_{Pl}^2 F(\phi,\chi) R - \frac{1}{2} \partial_\mu\phi \partial^\mu \phi  
-\left(V(\phi)-\frac{\alpha}{8}\chi^4\right)\Bigg\}
\end{eqnarray}
where  $\chi$ is the non-dynamical auxiliary field and
\begin{eqnarray}\label{conformal-factor_nq}
F(\phi,\chi)=\left(1+\frac{\xi}{M_{Pl}^2}\phi^2 -\frac{\alpha}{M_{Pl}^2}\chi^2\right).
\end{eqnarray}
By performing the conformal transformation $g_{\mu \nu}\rightarrow F(\phi,\chi)g_{\mu \nu}$, (\ref{symmergent-case_nm}) can be cast into the minimal form
\begin{eqnarray}\label{symmergent-case_m} 
S_{NQ}=\int d^4x \sqrt{-g} \Bigg\{
\frac{1}{2}M_{Pl}^2 R -\frac{1}{2} \left(\frac{\nabla \phi}{\sqrt{F(\phi,\chi)}}\right)^2  -\frac{V_J(\phi,\chi)}{F(\phi,\chi)^2}\Bigg\},
\end{eqnarray}
where $V_J(\phi,\chi)=V(\phi)-\alpha \chi^4/8$.

Variation of (\ref{symmergent-case_m}) with respect to $\chi$ and then using the least action principle ($\delta_{\chi} S_{NQ}=0$) allow for solving the auxiliary field $\chi$ in terms of the inflaton $\phi$

\begin{eqnarray}\label{symmergent-auxiliary}
\chi^2=\frac{4V(\phi)+\left(1+\frac{\xi}{M_{Pl}^2}\phi^2\right)(\nabla \phi)^2}{\left(\frac{1}{2}M_{Pl}^2 + \frac{1}{2}\xi\phi^2\right)+ \frac{\alpha}{M_{Pl}^2}(\nabla \phi)^2}.
\end{eqnarray}
Then applying slow-roll approximation $(\nabla \phi)^2 \ll \phi^2 $, takes (\ref{symmergent-auxiliary}) into
\begin{eqnarray}\label{symmergent-auxiliary_appr}
\chi^2=\frac{8V(\phi)}{M_{Pl}^2\left(1+\frac{\xi}{M_{Pl}^2}\phi^2\right)},
\end{eqnarray}
Substituting (\ref{symmergent-auxiliary_appr}) into  (\ref{symmergent-case_m}) takes us to, 

\begin{eqnarray}\label{symmergent-case_m_EF}
S_{NQ}=\int d^4x \sqrt{-g} \Bigg\{
\frac{1}{2}M_{Pl}^2 R -\frac{1}{2} \left(\frac{\nabla \phi}{\sqrt{Z_3(\phi)}}\right)^2  -\frac{V_J(\phi)}{Z_3(\phi)^2}\Bigg\},
\end{eqnarray}
where 
\begin{eqnarray} \label{VJ_EF}
V_J(\phi)=V(\phi)- 8 \alpha\left[\frac{V(\phi)}{M_{Pl}^2\left(1+\frac{\xi}{M_{Pl}^2}\phi^2\right)}\right]^2,
\end{eqnarray}
and 
\begin{eqnarray} \label{f_EF}
Z_3(\phi)=1+\frac{\xi}{M_{Pl}^2}\phi^2-\frac{8\alpha V(\phi)}{M_{Pl}^4 \left(1+\frac{\xi}{M_{Pl}^2}\phi^2\right)}.
\end{eqnarray}
Then, the resulting Einstein frame potential $V_{NQ}(\phi)=V_J(\phi)/Z_{3}(\phi)^2$ is

\begin{eqnarray} \label{V_EF}
V_{NQ}(\phi)=\frac{V(\phi)-\frac{8\alpha V(\phi)^2}{M_{Pl}^4 \left(1+\frac{\xi}{M_{Pl}^2}\phi^2\right)^2}}{Z_{3}(\phi)^2}\;,
\end{eqnarray}
where $V(\phi)$ is the reparametrized potential given in (\ref{pot_rep}).

This concludes the theoretical details regarding the inflationary potentials that we will analyze below. It is also an appropriate point to discuss the frame independence of cosmological perturbations. Within the field, it is common practice to convert the action of scalar-tensor theory from the Jordan frame to the Einstein frame (EF) to reduce it to its minimal form and calculate the inflationary observables using the EF potential. The validity of this approach fundamentally relies on the equivalence of the power spectra of cosmological perturbations between both frames. While this equivalence is well-established in the metric formalism, it has received less attention in the Palatini formulation.

The paper  \cite{Kubota_2021} by Kubota et al. investigates this equivalence within the context of the Palatini formalism. In this study, the authors derive quadratic actions for cosmological scalar and tensor perturbations in each frame and analyze the corresponding sound speeds. The calculations indicate that the quadratic actions are identical in both frames, leading to the conclusion that the observables are also the same. This investigation justifies the choice of the Einstein frame as the most convenient frame for conducting the numerical analysis of the parameter space of inflationary observables.  

 In the next section, we will continue with definitions of some observables pertinent to inflation and the methodology that we will use to scan the parameter spaces through numerical analyses.


\section{Inflationary observables and model parameters} \label{Sec.4}

In the ensuing sections, we will utilize the slow-roll (SR) approximation of inflationary dynamics to scan the parameter space of each of the model lagrangians given in the previous section. The analyses will be done in Planck units where $M_{Pl}$ is set to unity. The parameters that are specific to slow-roll, SR parameters for short, are usually defined through various derivatives of the Einstein frame potential written in terms of the canonical scalar field  ($\sigma$). Their traditional definitions are   \cite{Lyth:2009zz}
\begin{equation}\label{slowroll1}
\epsilon =\frac{1}{2}\left( \frac{V_{\sigma} }{V}\right) ^{2}\,, \quad
\eta = \frac{V_{\sigma\sigma} }{V}  \,, \quad
\zeta ^{2} = \frac{V_{\sigma} V_{\sigma\sigma\sigma} }{V^{2}}\,,
\end{equation}
where the subscript ($\sigma$) indicates derivative with respect to the canonical field. The definitions of the inflationary observables in terms of the SR parameters are given as 
\begin{eqnarray}\label{nsralpha1}
n_s = 1 - 6 \epsilon + 2 \eta \,,\quad
r = 16 \epsilon, \nonumber\\
\alpha_{s} \equiv \frac{\mathrm{d}n_s}{\mathrm{d}\ln k} = 16 \epsilon \eta - 24 \epsilon^2 - 2 \zeta^2\ ,
\end{eqnarray}
where $n_s$ is the (scalar) spectral index, $r$ is the tensor-to-scalar ratio and $\alpha_{s}$ is the running of the spectral index. 

Recently, a more accurate set of constraints on these inflationary observables are provided by the BICEP/Keck \cite{BICEP:2021xfz}, particularly on the tensor-to-scalar ratio, which tightens it to $r < 0.036$ at $95\%$ confidence level (CL). This robust constraint allows a sound explanation for not only the amplitude of the primordial gravitational waves but also the inflationary scale. In addition to the tensor-to-scalar ratio, recent BICEP/Keck results constrain the spectral index  $n_s$, to the range $[0.957, 0.976]$ at $2\sigma$ CL. These constraints are given for the pivot scale $k_* = 0.002$ Mpc$^{-1}$. 

Yet another restraint comes from the Planck 2018 measurements together with the baryon acoustic oscillations (BAO) results. They supply the constraint on $\alpha_s$ parameter: $\alpha_{s} = -0.0041 \pm 0.0067$ to base $\Lambda$CDM in $68\%$, TT,TE,EE +lowE+lensing+BAO \cite{Aghanim:2018eyx}. It should be noted that although the most recent constraints on the $\alpha_{s}$ are not very strong to test the existing inflationary models precisely, some improvements are expected with the observations of the 21-cm line in the near future \cite{Kohri:2013mxa,Basse:2014qqa,Munoz:2016owz}. 

For $r > 0.003$, at larger than $5\sigma$, primordial gravitational waves will be detectable by future CMB-S4 \cite{Abazajian:2019eic}. The highest limit of $r < 0.001$ at $95\%$ CL that CMB-S4 may reach, even in the absence of a detection, would provide important new insights into inflation \cite{Abazajian:2019eic}. It is anticipated that the CMB-S4 results may place strong constraints on inflationary models. The new results from CMB-S4 can even rule out most existing models in the future. More importantly, the robust constraints from CMB-S4 will significantly shape the perspective on the theory in the coming years. 

Returning to the definitions of inflationary observables; the number of e-folds of inflation that took place after the CMB pivot scale $k_{*}$ first exited the horizon is another significant quantity. In the SR approximation, it is calculated using 
\begin{equation} \label{efold1}
N_*=\int^{\sigma_*}_{\sigma_e}\frac{V\rm{d}\sigma}{V_{\sigma}} \: ,
\end{equation}
where the subscripts “$*$” and “$e$” mark the moment when the pivot scale exits the horizon and the end of inflation, respectively. The end of SR inflation is determined by  $\epsilon(\sigma_e) = 1$. 

The horizon problem can be solved if $N_*$ is around $60$. The precise value of $N_*$  depends on how the universe has evolved. We assume a standard thermal history after inflation \cite{Liddle:2003as} and consider $N_*$ to have the following form
\begin{eqnarray} \label{efoldsreal}
N_*\approx64.7+\frac12\ln\rho_*-\frac{1}{3(1+\omega_{reh})}\ln\rho_e
+\Big(\frac{1}{3(1+\omega_{reh})}-\frac14\Big)\ln\rho_{reh},
\end{eqnarray} 
where the energy densities at the end of inflation and the end of reheating are $\rho_{e}=(3/2)V(\phi_{e})$ and $\rho_{reh}$, respectively. The standard model value $g_*=106.75$, which represents the number of relativistic degrees of freedom, can be used to find $\rho_{reh}$. Furthermore, at the pivot scale, the energy density is $\rho_{*} \approx V(\phi_*)$.

The definitions of $\rho_{reh}$ and $\rho_*$ are given in the following forms
\begin{equation}
 \rho_{reh} = \Big(\frac{\pi^2}{30}106.75\Big) T_{reh}^4,  \qquad  \rho_{*} = \frac{3 \pi^2\Delta^2_\mathcal{R} r }{2},
\end{equation}
where, $T_{reh}$ corresponds to the reheat temperature, $r$ is the tensor-to-scalar ratio and $\Delta^2_\mathcal{R}$ is the amplitude of curvature perturbation defined by
\begin{equation} \label{perturb1}
\Delta_\mathcal{R}^2=\frac{1}{12\pi^2}\frac{V^3}{V_{\sigma}^2}\; .
\end{equation}
For the pivot scale $k_* = 0.002$ Mpc$^{-1}$, the amplitude of curvature perturbation should be $\Delta_\mathcal{R}^2\approx 2.1\times10^{-9}$ to comply with the Planck results \cite{Aghanim:2018eyx}. Lastly, $\omega_{reh}$ in (\ref{efoldsreal}) is the equation of state parameter during reheating. In this work, we take $\omega_{reh}=1/3$, which is equivalent to assuming instant reheating. Under instant reheating assumption $N_*$ in (\ref{efoldsreal}) turns into
\begin{eqnarray} \label{efoldsreal2} 
N_*\approx64.7+\frac12\ln\rho_*-\frac{1}{4}\ln\rho_e \:.
\end{eqnarray}
Equation (\ref{efoldsreal2}) reminds us that inflationary predictions should not depend at all on $T_{reh}$ in the case of instant reheating.

Although we have defined the SR parameters in terms of the canonical field $\sigma$ in (\ref{slowroll1}), it is more convenient to rewrite them in terms of the original scalar field, $\phi$, for our numerical computations. We require this redefinition because for general values of the free parameters, it is not easy, if not impossible, to express the inflationary potential in terms of the canonical field. 

It is possible to switch between the two definitions of the SR parameters through the use of conformal factors $Z_i (i=1,2,3)$ defined for each case in Sec. \ref{Sec.3}. In terms of $\phi$; the definitions of SR parameters are as follows \cite{Linde:2011nh} 

\begin{eqnarray}\label{slowroll2}  
\epsilon=Z_i\epsilon_{\phi}\,,\quad
\eta=Z_i\eta_{\phi}+{\rm sgn}(V')Z_i'\sqrt{\frac{\epsilon_{\phi}}{2}}, \nonumber \\
\zeta^2=Z_i\left(Z_i\zeta^2_{\phi}+3{\rm sgn}(V')Z_i'\eta_{\phi}\sqrt{\frac{\epsilon_{\phi}}{2}}+Z_i''\epsilon_{\phi}\right),
\end{eqnarray}
where, 
\begin{equation}
\epsilon_{\phi} =\frac{1}{2}\left( \frac{V^{\prime} }{V}\right) ^{2}\,, \quad
\eta_{\phi} = \frac{V^{\prime \prime} }{V}  \,, \quad
\zeta ^{2} _{\phi}= \frac{V^{\prime} V^{\prime \prime\prime} }{V^{2}}\,
\end{equation}
are defined with the derivatives taken with respect to $\phi$. Likewise, the number of e-folds \eqref{efold1} and the amplitude of curvature perturbations \eqref{perturb1} can be written in terms of $\phi$ as
\begin{eqnarray}\label{perturb2}
N_*&=&\rm{sgn}(V')\int^{\phi_*}_{\phi_e}\frac{\mathrm{d}\phi}{ Z_i(\phi)\sqrt{2\epsilon_{\phi}}}\,,\\
\label{deltaR} \Delta_\mathcal{R}&=&\frac{1}{2\sqrt{3}\pi}\frac{V^{3/2}}{\sqrt{Z_i}|V^{\prime}|}\,.
\end{eqnarray}

In the following sections, we present our numerical results for the inflationary predictions of the large field polynomial inflation potential in terms of $\phi$. Since instant reheating has been assumed, the results will be independent of reheat temperature in any case. 

\begin{figure}[!htbp]
\centering
    \includegraphics[width=0.445\textwidth]{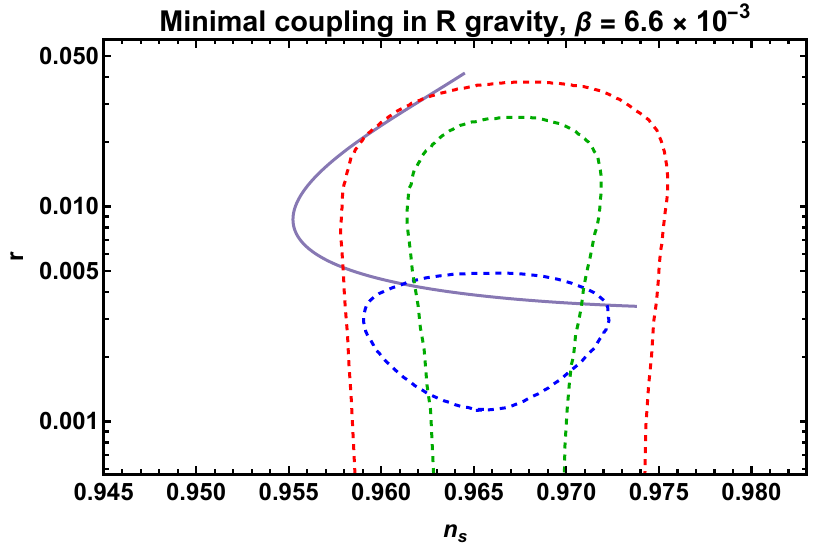}
    \includegraphics[width=0.445\textwidth]{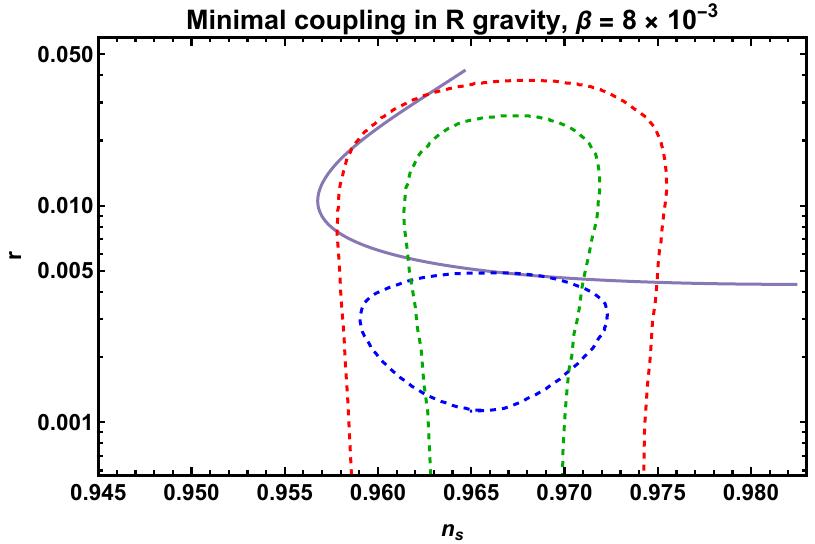}
    
    \
    
    \includegraphics[width=0.91\textwidth]{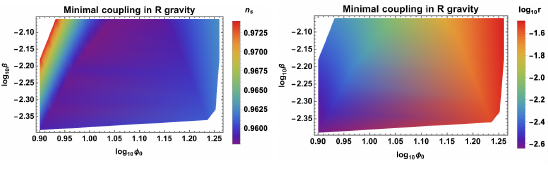}
    \caption{\label{fig:1} Top panel displays the $n_s-r$ predictions for the polynomial inflation with minimal coupling in $R$ gravity for $\beta = 6.6 \times 10^{-3}$ (left pane) and $\beta = 8 \times 10^{-3}$ (right pane). Red (green) dashed contours indicate the recent $95\% \: (68\%)$ CL given by BICEP/Keck \cite{BICEP:2021xfz} and the blue dashed contours correspond to the sensitivity forecast for the future CMB-S4 \cite{Abazajian:2019eic}. The bottom panel displays the $\beta-\phi_0$ planes for which $n_s$ (left pane), $r$ (right pane) predictions are within the recent $95\% \: (68\%)$ BICEP/Keck CL contours.}
\end{figure}

\subsection{Predictions in metric-Palatini \boldmath{$R$} gravity} \label{subsec.4.1}
\subsubsection{Minimal coupling} \label{subsubsec.4.1.1}

In this section, we discuss the numerical results for polynomial inflation with minimal coupling in $R$ gravity. The inflaton potential that drives the expansion of the universe in this case is given in (\ref{pot_rep}). Numerical analysis has been conducted for the inflationary parameters such that the inflaton values satisfy the conditions of polynomial inflation together with $\phi \lesssim \phi_0$. This way, it has been ensured that the inflationary excursion takes place over the concave portion of the potential profile in the vicinity of the inflection point $\phi_0$.

Polynomial inflation, in the metric formalism, has been analyzed for a range of the number of e-folds, specifically for the range of $50-65$, in \cite{Drees:2022aea} previously. Unlike the approach taken there, in this work, the analysis is based on a method that involves two different ways of calculating the e-folds parameter. The evolution of the universe is taken into account through the use of (\ref{efoldsreal2}) in addition to the usual SR way of calculating $N_*$ via (\ref{perturb2}). Note that (\ref{efoldsreal2}) assumes instant reheating for the evolution of the universe. The inflationary parameters, $n_s, r$, and $\alpha_s$ are computed following an iteration algorithm crudely based on the parameter $d$ in the potential. The finer iterations are carried on over the other variables of the model until the two different ways of computing the e-folds parameter converge.

\begin{figure}[!htbp]
\centering
     \includegraphics[width=0.45\textwidth]{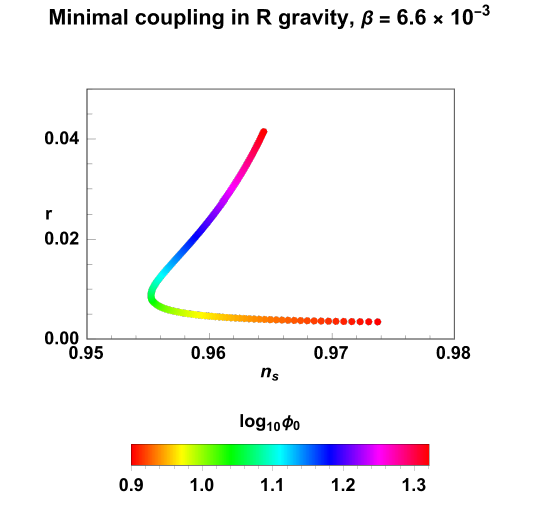}
     \includegraphics[width=0.45\textwidth]{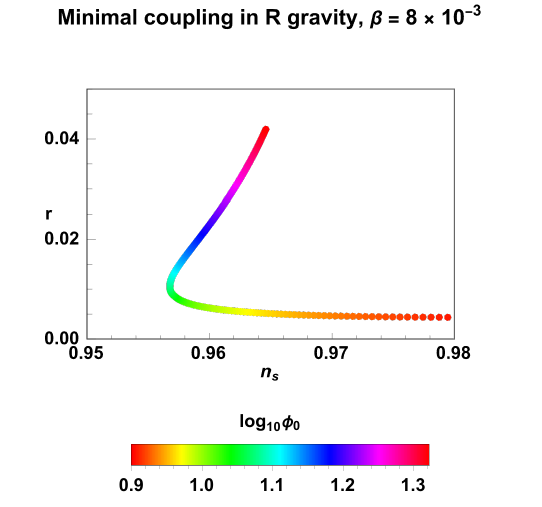}
    \caption{\label{fig:2}  Minimal coupling in $R$ gravity. The relationship between  the $n_s-r$ vs $\phi_ 0$ for $\beta = 6.6 \times 10^{-3}$ (left pane) and $\beta = 8 \times 10^{-3}$ (right pane).   
    }
\end{figure}

\begin{table}[!htbp]
\centering
\begin{tabular}{|c|c|c|c|c|c|c|}
\hline
$\phi_0$ & $\phi_{*}$ & $d/10^{-14}$ & $N_{*}$ & $n_s$ & $r/ 10^{-3}$ & $\alpha_s / 10^{-4}$ \\
\hline
7.94   & 7.5406    & 7.71 & 59.4    & 0.9737 & 3.44  & -16.81 \\
8.56   & 7.9306    & 6.48 & 59.5    & 0.9650 & 3.89  & -13.50 \\
9.70   & 8.6102    & 5.49 & 59.6    & 0.9575 & 5.38  & -10.04 \\
10.30  & 8.9431    & 5.22 & 59.7    & 0.9560 & 6.47  & -8.89 \\
11.91  & 9.7456    & 4.72 & 59.8    & 0.9554 & 10.27 & -6.92 \\
12.46  & 9.9902    & 4.58 & 59.9    & 0.9558 & 11.83 & -6.48 \\
13.84  & 10.5451   & 4.21 & 60.0    & 0.9573 & 16.19 & -5.70 \\
14.69  & 10.8493   & 3.98 & 60.1    & 0.9584 & 19.14 & -5.39 \\
15.76  & 11.1902   & 3.69 & 60.1    & 0.9597 & 22.96 & -5.13 \\
16.16  & 11.3079   & 3.58 & 60.2    & 0.9602 & 24.42 & -5.06 \\
17.95  & 11.7776   & 3.11 & 60.3    & 0.9621 & 30.96 & -4.87 \\
18.31  & 11.8624   & 3.02 & 60.3    & 0.9624 & 32.28 & -4.85 \\
19.25  & 12.0678   & 2.80 & 60.3    & 0.9632 & 35.62 & -4.81 \\
20.95  & 12.3951   & 2.44 & 60.4    & 0.9644 & 41.46 & -4.79 \\ 
\hline
\end{tabular}
\caption{Sample set of model parameters and corresponding predictions for the minimal coupling in $R$ gravity with $\beta = 6.6 \times 10^{-3}$.}
\label{table:minimal}
\end{table}
Figures \ref{fig:1} and \ref{fig:2} give a summary of the parameter space that satisfies inflation in the minimal case. The results demonstrate that the $n_s-r$ predictions fall within the $68\%$ CL contour of recent BICEP/Keck data for $\beta = 6.6 \times 10^{-3}$ while $\phi_0$ resides in the range $[8.12,\: 8.90]$. Similarly, for $\beta = 8 \times 10^{-3}$ and $\phi_0 \sim [8.57,\: 9.54]$ the $68\%$ CL mark is also attained. The $n_s-r$ predictions hit the $95\%$ CL of BICEP/Keck data for $\beta = 6.6 \times 10^{-3}$ with $\phi_0 \sim [8.91,\: 9.6]$, and for $\beta = 8 \times 10^{-3}$ with $\phi_0 \sim [9.55,\: 10.6]$.

Also, for $\beta = 8 \times 10^{-3}$ with $\phi_0 \sim [10.6,\: 14.3]$, the parameters are outside the observational region, and the predictions can be within the $95\%$ CL for $\phi_0 \sim [14.31,\: 17.59]$ but for $\phi_0 \gtrsim 17.6$, the inflationary parameters cannot be inside the observational region at all. As for $\beta = 6.6 \times 10^{-3}$, with $\phi_0 \sim [9.6,\: 15.0]$ the parameters are outside the observational region, and the predictions can be inside the $95\%$ CL for $\phi_0 \sim [15.1,\: 17.7]$ but for $\phi_0 \gtrsim 17.6$, the inflationary parameters cannot be inside the observational region. Moreover, for $\phi_0 \sim 7.94$, $n_s-r$ predictions remain inside the $95\%$ confidence level for $\beta = 6.6 \times 10^{-3}$, but ruled out for $\beta = 8 \times 10^{-3}$.

We have also compared our findings with the anticipated sensitivity forecast for CMB-S4. Our analysis indicates that the $n_s-r$ predictions for $\beta = 6.6 \times 10^{-3}$ fall within the expected CMB-S4 sensitivity region for lower $\phi_0$ values, around $\phi_0 \sim [8.08,\: 8.90]$. However, for $\beta = 8 \times 10^{-3}$, our results do not align with the projected CMB-S4 forecast.

As shown in the bottom panel of Figure \ref{fig:1}, the inflationary predictions fall within the observational region for the following ranges of $\beta$ and $\phi_0$: $3.9 \times 10^{-3} \lesssim \beta \lesssim 9.3 \times 10^{-3}$ and $7.94 \lesssim \phi_0 \lesssim 17.8$. For the specified values of $\beta$ and $\phi_0$, the inflationary parameters have been determined to lie within the ranges of $0.959 \lesssim n_s \lesssim 0.973$ and $2.5 \times 10^{-3} \lesssim r \lesssim 3.2 \times 10^{-2}$. 
For instance, when $\phi_0 \approx 8.51$ and $\beta \approx 7 \times 10^{-3}$, the predicted values are $n_s \approx 0.9725$ and $r \approx 0.005$. Similarly, for $\phi_0 \approx 15.84$ and $\beta \approx 5 \times 10^{-3}$, the predicted values are $n_s \approx 0.962$ and $r \approx 2.5 \times 10^{-3}$.

It should be noted that, as $\phi_0$ experiences an increase, there is a corresponding increase in the tensor-to-scalar ratio $r$. This can be observed in the rainbow graph of the bottom right pane of Figure \ref{fig:1} and both panes of Figure \ref{fig:2}. A similar relationship exists between the $\beta$ and $n_s$ for lower $\phi_0$. The left pane of the bottom panel of Figure \ref{fig:1} manifests that as $\beta$ gets larger the spectral index $n_s$ also builds up. A quick comparison between the left and right panes of Figure \ref{fig:2} reveals the same relationship.  

The relationship of $\phi_0$ with $n_s$ is more complicated than that of its relation with $r$. As $\phi_0$ increases, it is evident from Figure \ref{fig:2} and Table \ref{table:minimal} that the spectral index $n_s$ initially decreases, reaches a minimum, and then starts to increase again. The turning point appears to be situated around $\phi_0 \approx 12$.

By examining the right pane of the bottom panel in Figure \ref{fig:1}, we can see that the relationship between $\beta$ and $r$ varies depending on the position along the $\phi_0$ axis. For example, when $\phi_0$ is at the far left, increasing $\beta$ has minimal impact on the tensor-to-scalar ratio. However, as $\phi_0$ increases, we start to notice that a rise in $\beta$ leads to an increase in $r$.

Table \ref{table:minimal} displays the various model parameters for the e-folds in the range of approximately  $59-60$. While it may not be receiving much attention at the moment, the value of $ | \alpha_s| $ is found to be in the range of $(10^{-4}, 10^{-3})$.

At this point, the results presented above can be compared with those of \cite{Drees:2022aea} by choosing similar parameter values. It is reported in \cite{Drees:2022aea} that for $\beta = 8\times 10^{-3}$, $\phi_0 = 17$ and $\phi_*=11.61$, the expected values for inflation are $n_s \sim 0.9619$, $r\sim 2.7 \times 10^{-2}$ and $N_*\sim 61.1$. The inflationary observables in this study for the parameter selection $\beta = 6.6 \times 10^{-3}$, $\phi_0= 17.95$ and $\phi_*=11.77$ which are close to that of \cite{Drees:2022aea} are computed to be $n_s \sim 0.9621$, $r\sim 3 \times 10^{-2}$ and $N_*\sim 60.3$.  It is important to highlight that our findings are generally consistent with those in \cite{Drees:2022aea}, including the value of $|\alpha_s|$, which has been determined to be on the order of $(10^{-4}, 10^{-3})$.

In addition, the results of this study indicate that as $\beta$ increases, the value of $n_s$ also increases, as shown in the bottom panel of Figure \ref{fig:1}. Furthermore, the $r$ values decrease as $\phi_0$ decreases, consistent with the findings in reference \cite{Zhang:2024ldx}.
In light of all these observations, it is safe to say that the outcomes of the numerical analysis conducted in this section align with the existing literature \cite{Drees:2022aea} and \cite{Zhang:2024ldx}. This situation is expected since, as it has already been asserted at the end of Sec. \ref{subsubsec. 3.1.1}, the Palatini formulation of the minimal coupling to $R$ gravity is equivalent to the metric formulation at the equation of motion level. 

The discussion in this section serves as an important initial step in the exploration of polynomial inflation in Palatini $f(R,\phi)$ gravity. In the upcoming sections, we will expand our analysis to include the examination of the non-minimal coupling in $R$ gravity, as well as both minimal and non-minimal coupling in $f(R,\phi)$ gravity.

\begin{figure}[!htbp]
\centering
    \includegraphics[width=0.48\textwidth]{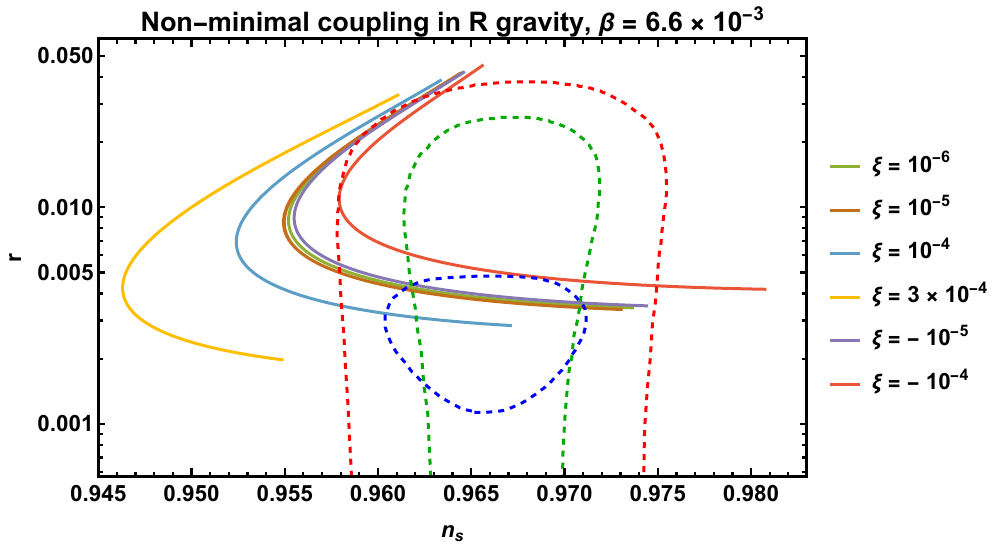}
    \includegraphics[width=0.48\textwidth]{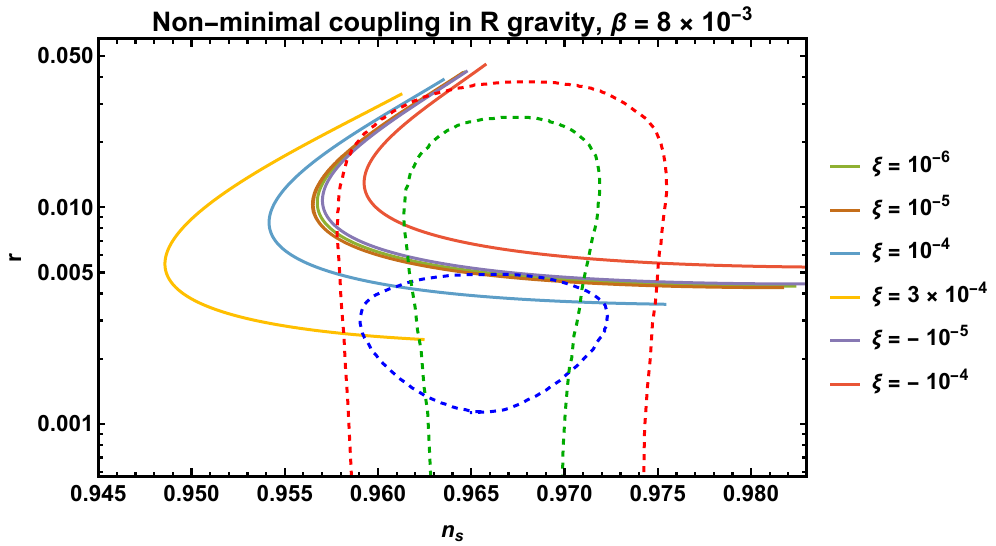}
    
    \  
    
    \includegraphics[width=0.88\textwidth]{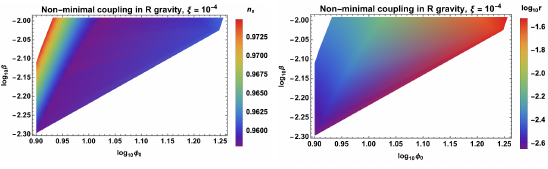}
    \caption{\label{fig:3} Top panel shows the $n_s-r$ predictions for the polynomial inflation with non-minimal coupling in $R$ gravity for $\beta = 6.6 \times 10^{-3}$ (left pane) and $\beta = 8 \times 10^{-3}$ (right pane) for selected $\xi$ values. The red (green) dashed contours indicate the recent $95\% \: (68\%)$ CL given by BICEP/Keck and the blue dashed contours correspond to the sensitivity forecast for the future CMB-S4. The bottom panel shows the $\beta-\phi_0$ planes for which $n_s$ (left pane), $r$ (right pane) predictions are within the recent $95\%\:(68\%)$ BICEP/Keck CL contours for $\xi=10^{-4}$.}
\end{figure}

\begin{figure}[tbp]
\centering
    \includegraphics[width=0.45\textwidth]{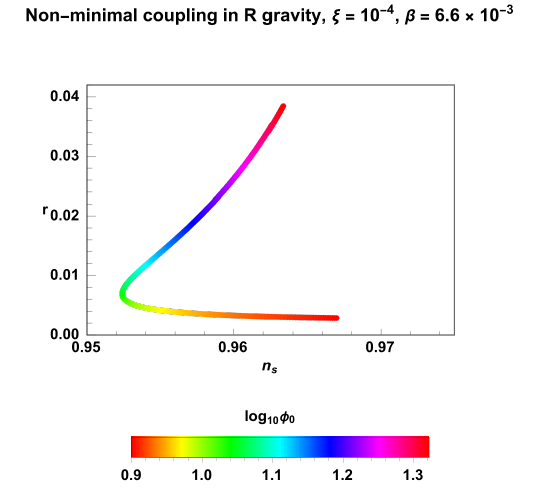}
    \caption{\label{fig:4} Non-minimal coupling in $R$ gravity. The relationship between  the $n_s-r$ vs $\phi_ 0$ for $\beta = 6.6 \times 10^{-3}$ and $\xi = 10^{-4}$.
    }
\end{figure}

\subsubsection{Non-minimal coupling} \label{subsubsec.4.1.2}

In this section, the analysis of the PI model has been expanded to include the non-minimal coupling of inflaton in $R$ gravity. The details of the model are given in Sec. \ref{subsubsec. 3.1.2}. For example, the action is presented in equation (\ref{nonminimal-case}), and the potential is described in equation (\ref{nonminimal-case_pot_EF}).  The numerical analysis performed is summarized in Figures \ref{fig:3} and \ref{fig:4}, along with Table \ref{table:nonminimal_linear}.

In the top panel of Figure \ref{fig:3}, we present the results for selected values of the non-minimal coupling parameter, $\xi$. We find that the non-minimal coupling values that meet observational constraints are very small. According to \cite{Gialamas:2023flv}, in the context of $R$-gravity, both metric and Palatini formalisms should yield results that are largely indistinguishable from an observational standpoint when using such small non-minimal couplings. The exploration of non-minimally coupled polynomial inflation using the metric formalism is still an uncharted area. Therefore, a future investigation of non-minimally coupled polynomial inflation within the metric framework is expected to produce results comparable to the Palatini ones within the small non-minimal coupling regime.

Inspecting both panes of top panel of Figure \ref{fig:3} reveals the general effect of  $\xi$ on $n_s  -  r$. It appears that an increase in the non-minimal coupling $\xi$ causes a decrease in both $n_s$ and $r$.  When examining either $\beta = 6.6 \times 10^{-3}$ or $\beta = 8 \times 10^{-3}$ pane, it is evident that the $n_s  -  r$ curves converge to the respective ones drawn in the top panel of Figure \ref{fig:1} as the limit $\xi \rightarrow 0$ is approached. 
For instance, when looking at the curves for $\xi = 10^{-6}$, $\xi = 10^{-5}$, and $\xi = -10^{-5}$, it becomes apparent that they exhibit minimal deviation from each other, indicating a high level of similarity.
Additionally, for these $\xi$ values, the inflationary predictions are ruled out for smaller $\phi_0$ values, approximately $\phi_0 \sim 7.95$ for $\beta = 8 \times 10^{-3}$, because $n_s \sim 0.98$, which is outside the observational ranges. However, the inflationary predictions are within the $2\sigma$ CL for $\beta = 6.6 \times 10^{-3}$  and $\phi_0$ $\sim 7.95$, with $n_s \sim 0.973$ and $r \sim 4 \times 10^{-3}$.

\begin{table}[!htbp]
\centering
\begin{tabular}{|c|c|c|c|c|c|c|}
\hline
$\phi_0$ & $\phi_{*}$ & $d/10^{-14}$ & $N_{*}$ & $n_s$ & $r/ 10^{-3}$ & $\alpha_s / 10^{-4}$ \\
\hline
7.94  & 7.5397  & 7.70 & 59.4 & 0.9736 & 3.43 & -16.79 \\
8.38  & 7.8189  & 6.74 & 59.4 & 0.9669 & 3.72 & -14.28 \\
9.49  & 8.4881  & 5.61 & 59.6 & 0.9583 & 5.04 & -10.52 \\
10.24 & 8.9089  & 5.24 & 59.6 & 0.9561 & 6.34 & -8.99 \\
11.34 & 9.4753  & 4.88 & 59.8 & 0.9552 & 8.77 & -7.48 \\
12.20 & 9.8757  & 4.64 & 59.9 & 0.9556 & 11.07 & -6.67 \\
13.44 & 10.3936 & 4.31 & 60.0 & 0.9568 & 14.87 & -5.88 \\
14.65 & 10.8335 & 3.99 & 60.1 & 0.9583 & 18.97 & -5.40 \\
15.28 & 11.0403 & 3.81 & 60.1 & 0.9591 & 21.20 & -5.23 \\
16.57 & 11.4224 & 3.47 & 60.2 & 0.9606 & 25.90 & -5.00 \\
17.41 & 11.6466 & 3.25 & 60.2 & 0.9615 & 29.01 & -4.91 \\
18.18 & 11.8319 & 3.05 & 60.3 & 0.9623 & 31.79 & -4.85 \\
19.77 & 12.1737 & 2.69 & 60.3 & 0.9636 & 37.43 & -4.80 \\
20.29 & 12.2740 & 2.57 & 60.3 & 0.9640 & 39.22 & -4.79 \\
\hline
\end{tabular}
\caption{Sample set of model parameters and corresponding predictions for the non-minimal coupling in $R$ gravity with $\beta = 6.6 \times 10^{-3}$ and $\xi=10^{-4}$.}
\label{table:nonminimal_linear}
\end{table}

For $\xi = 3 \times 10^{-4}$, the predictions do not fall within the observational boundary for $\beta = 6.6 \times 10^{-3}$. However, for $\beta = 8 \times 10^{-3}$, the predictions can even fall within the $1\sigma$ confidence level for $\phi_0 \sim 7.94$, with $n_s \sim 0.9631$ and $r \sim 2.4 \times 10^{-3}$.

For $\xi=-10^{-4}$ at the negative far end, the $n_s-r$ predictions are ruled out for smaller $\phi_0$, approximately $ \phi_0 \sim 8$, for both $\beta = 6.6 \times 10^{-3}$ and $\beta = 8 \times 10^{-3}$ values. Conversely, with an increase in $\phi_0$, the predictions can align with the cosmological data. For example, with $\xi = - 10^{-4}$ and $\beta = 6.6 \times 10^{-3}$,
we obtain $N_* \sim 59.5$, $n_s \sim 0.9662$, and $r \sim 0.0051$ for $\phi_0 \sim 8.965$. Similarly, for $\xi = - 10^{-4}$ and $\beta = 8 \times 10^{-3}$, at $\phi_0 \sim 8.965$, we find $N_* \sim 59.5$, $n_s \sim 0.9711$, and $r \sim 5.8 \times 10^{-3}$.

Let's consider another example where we take $\xi = 10^{-4}$. In this case the $n_s-r$ can be within $1 \sigma$ CL for $\beta = 6.6 \times 10^{-3}$ and $\phi_0$ $\sim 7.94$. However, for the same $\phi_0$ the predictions are ruled out for $\beta = 8 \times 10^{-3}$. In addition, for $\phi_0 \sim 9$ and $\beta = 8 \times 10^{-3}$, the inflationary predictions can enter both $1 \sigma$ CL for BICEP/Keck data and future CMB-S4 sensitivity forecast region.

It's worth noting that, unlike the minimal coupling case, in $R$ gravity with non-minimal coupling, extended regions of $\beta-\phi_0$ exist where the predictions for $n_s$ and $r$ fall outside the observational range. This is illustrated in the bottom panel of Figure \ref{fig:3}. As $\phi_0$ increases, these regions become more prominent. For example, when $\phi_0 \sim 12.58$ and $\beta \sim 5.6 \times 10^{-3}$, or when $\phi_0 \sim 14.78$ and $\beta \sim 6.6 \times 10^{-3}$, the predictions for $n_s-r$  do not fall within the observable range, even though these values of $\phi_0$ and $\beta$ satisfy the conditions for successful inflation. It can be concluded that each $\beta$ value has a maximum $\phi_0$ value where the conditions of inflation are satisfied and remain consistent with recent cosmological data from BICEP/Keck. Unlike the minimal coupling, in this case, there is no smooth distribution that fits within the observational ranges for all $\beta-\phi_0$ values. As a result, the outcomes for non-minimal coupling differ slightly from those for the minimal one. Let us give some examples of our results within the $\beta-\phi_0$ plane. For instance, $(\phi_0, \beta) \sim (8.12, 7 \times 10^{-3})$ $\rightarrow$ $(n_s, r) \sim (0.9675, 3.16 \times 10^{-3})$. Similarly, $(\phi_0, \beta) \sim (14.12, 8 \times 10^{-3})$ $\rightarrow$ $(n_s, r) \sim (0.96, 2.5 \times 10^{-2})$. In addition, similar to the minimal case, we find that the predictions of $|\alpha_s|$ are in the order of $(10^{-4}, 10^{-3})$ for non-minimal coupling in $R$ gravity.

One last example can be drawn from Table \ref{table:nonminimal_linear}: for $\beta = 6.6 \times 10^{-3}$, $\xi=10^{-4}$, and $\phi_0 \sim 8.38$ we have  $n_s \sim 0.9669$, $r \sim 3.72 \times 10^{-3}$ and $|\alpha_s | \sim 1.428 \times 10^{-3}$. 

\begin{figure}[!htbp]
\centering
    \includegraphics[width=0.48\textwidth]{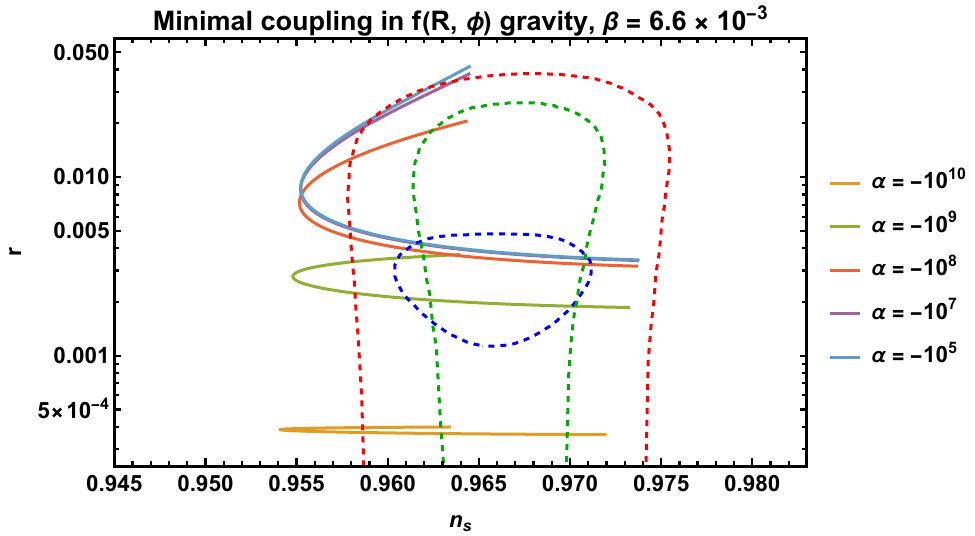}
    \includegraphics[width=0.48\textwidth]{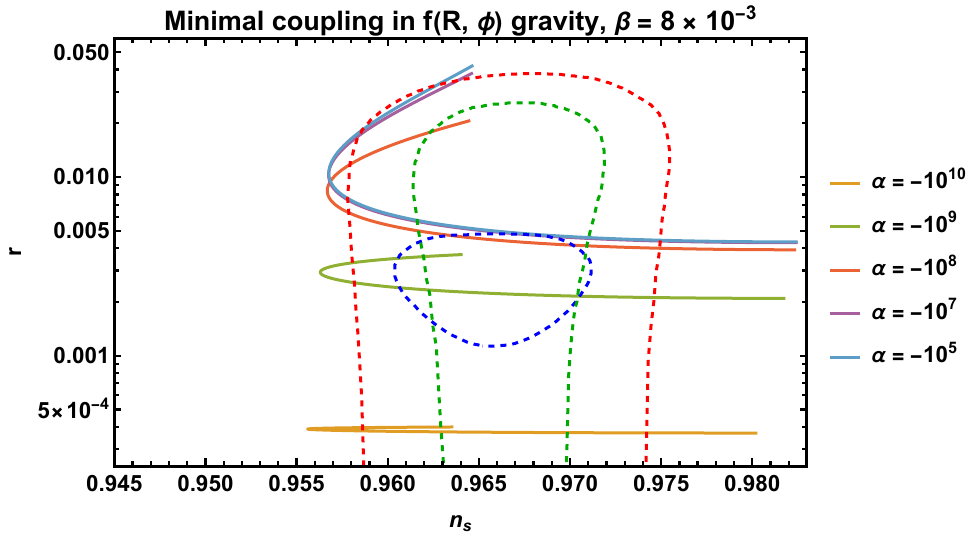}
    
    \
    
    \includegraphics[width=0.88\textwidth]{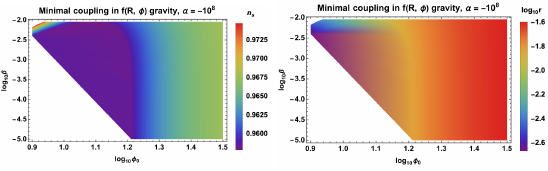}
    \caption{\label{fig:5} Top panel shows the $n_s-r$ predictions for the polynomial inflation with the minimal coupling in $f(R,\phi)$ gravity for $\beta = 6.6 \times 10^{-3}$ (left pane) and $\beta = 8 \times 10^{-3}$ (right pane) for selected $\alpha$ values. Red (green) dashed contours indicate the recent $95\% \: (68\%)$ CL given by BICEP/Keck \cite{BICEP:2021xfz} and the blue dashed contours correspond to the sensitivity forecast for the future CMB-S4 \cite{Abazajian:2019eic}. Bottom panel depicts the $\beta-\phi_0$ planes for which $n_s$ (left pane), $r$ (right pane) predictions are within the recent $95\% \:(68\%)$ BICEP/Keck CL contours and $\alpha=-10^{8}$. }
\end{figure}

\subsection{Predictions in metric-Palatini \boldmath{$f(R,\phi)$} gravity} \label{subsec.4.2}
\subsubsection{Minimal coupling} \label{subsubsec.4.2.1}

\begin{figure}[tbp]
\centering
    \includegraphics[width=0.45\textwidth]{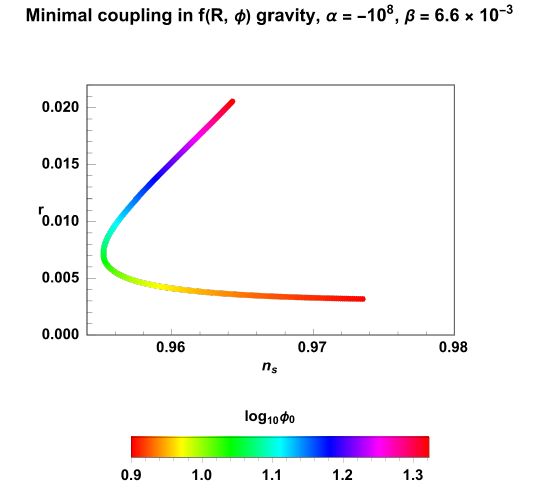}
    \caption{\label{fig:6} Minimal coupling in $f(R,\phi)$ gravity. The relationship between  the $n_s-r$ vs $\phi_ 0$ for $\beta = 6.6 \times 10^{-3}$ and $\alpha=-10^{8}$.
    }
\end{figure}

This section further analyzes the polynomial inflation within the Palatini formulation by incorporating the Starobinsky term. The action of the model is given in (\ref{minimal-quad-case}) and the potential in (\ref{V_MQ_EF_simp}). Figures \ref{fig:5} and \ref{fig:6} display the results of the numerical analysis for polynomial inflation with minimal coupling in $f(R,\phi)$ gravity. Additionally, Table \ref{table:minimal_quadratic} lists several model parameters along with the corresponding predictions for the case where $\beta = 6.6 \times 10^{-3}$ and  $\alpha = -10^{8}$.

The top panel of Figure \ref{fig:5} illustrates the predictions for the relationship between $n_s -r$ and $\alpha$ based on two specific values of the parameter $\beta$:  $6.6 \times 10^{-3}$ shown on the left and $8 \times 10^{-3}$ on the right. Within this framework, the parameter $\alpha$ plays a critical role by regulating the strength of the Starobinsky term, which is an essential component in cosmological models.
As we examine both panes, a clear trend emerges: increasing the absolute value of $\alpha$  facilitates a substantial reduction in the tensor-to-scalar ratio $r$. This reduction occurs while the spectral index $n_s$ and other related observables remain relatively stable. For instance, when we set $\alpha$ to a value of $-10^{10}$, we notice that the tensor-to-scalar ratio can decrease dramatically, reaching values as low as $r \sim 5 \times 10^{-4}$. This observation underscores the impact of $\alpha$ on the model's predictions regarding $r$ and it also aligns with the results conveyed in \cite{Antoniadis:2018ywb} and \cite{Enckell:2018hmo}.
\begin{table}[!htbp]
\centering
\begin{tabular}{|c|c|c|c|c|c|c|}
\hline
$\phi_0$ & $\phi_{*}$ & $d/10^{-14}$ & $N_{*}$ & $n_s$ & $r/ 10^{-3}$ & $\alpha_s / 10^{-4}$ \\
\hline
7.97  & 7.5549  & 7.65 & 59.4 & 0.9732 & 3.18 & -16.65 \\
8.50  & 7.8925  & 6.58 & 59.4 & 0.9655 & 3.50 & -13.76 \\
9.31  & 8.3831  & 5.75 & 59.5 & 0.9592 & 4.28 & -10.99 \\
10.47 & 9.0298  & 5.17 & 59.6 & 0.9557 & 5.83 & -8.64 \\
11.77 & 9.6761  & 4.78 & 59.7 & 0.9553 & 7.96 & -7.06 \\
12.49 & 10.0015 & 4.59 & 59.8 & 0.9558 & 9.24 & -6.47 \\
13.54 & 10.4241 & 4.32 & 59.8 & 0.9569 & 11.08 & -5.86 \\
14.97 & 10.9333 & 3.93 & 59.9 & 0.9586 & 13.49 & -5.34 \\
15.22 & 11.0111 & 3.87 & 59.9 & 0.9589 & 13.87 & -5.28 \\
16.37 & 11.3546 & 3.55 & 59.9 & 0.9603 & 15.57 & -5.06 \\
17.09 & 11.5480 & 3.37 & 60.0 & 0.9611 & 16.54 & -4.98 \\
18.49 & 11.8851 & 3.01 & 60.0 & 0.9624 & 18.20 & -4.88 \\
19.50 & 12.0978 & 2.78 & 60.0 & 0.9633 & 19.24 & -4.85 \\
20.17 & 12.2285 & 2.63 & 60.0 & 0.9638 & 19.86 & -4.85 \\
\hline
\end{tabular}
\caption{Sample set of model parameters and corresponding predictions for the minimal coupling in $f(R,\phi)$ gravity with $\beta = 6.6 \times 10^{-3}$ and $\alpha=-10^{8}$.}
\label{table:minimal_quadratic}
\end{table}

Inspection of the same panel reveals that the $n_s-r$ predictions satisfy both the current cosmological data and the anticipated future sensitivities for all selected values of $\alpha$. For instance, with $\alpha=-10^8$, we take $\beta \sim 5 \times 10^{-3}$ and $\phi_0 \sim 8.92$. This results in $n_s \sim 0.965$ and $r\sim 3\times 10^{-3}$, meaning that the inflationary predictions fall within the $1\sigma$ CL of the recent BICEP/Keck data.

Another point to mention is that similar to the $R$ gravity with non-minimal coupling, extended regions of the $\beta-\phi_0$ plane exist where the predictions for $n_s$ and $r$ fall outside the observational range. This is illustrated in the bottom panel of Figure \ref{fig:5}. However, in this case, these regions occupy the lower $\phi_0$ regime with $\phi_0 \lesssim 16.6$. This means these regions become less prominent as $\phi_0$ increases. Each $\phi_0$ with $\phi_0 \lesssim 16.6$ has a minimum $\beta$ for which the results are consistent with the recent BICEP/Keck data.

Let us recite  one of our results concerning the $\beta - \phi_0$ plane: with $\phi_0 \sim 12.5893$ and $\beta \sim 10^{-4}$, we find that $n_s \sim 0.959$ and $r \sim 1.26 \times 10^{-2}$.  Lastly, as it can be checked in the Table \ref{table:minimal_quadratic} with $\beta = 6.6 \times 10^{-3}$, $\alpha=-10^{8}$ and $\phi_0 \sim 19.5$ we find $N_* \sim 60$, $n_s \sim 0.963$ and $r \sim 1.98 \times 10^{-2}$ for the minimal coupling in $f(R,\phi)$ gravity. From this table, it is evident that the predictions of $|\mathrm{d}n_s / \mathrm{d}\ln k| $ are too small to be tested with the current cosmological data, similar to the case of minimal coupling in $R$ gravity. 
\begin{figure}[!htbp]
\centering
    \includegraphics[width=0.48\textwidth]{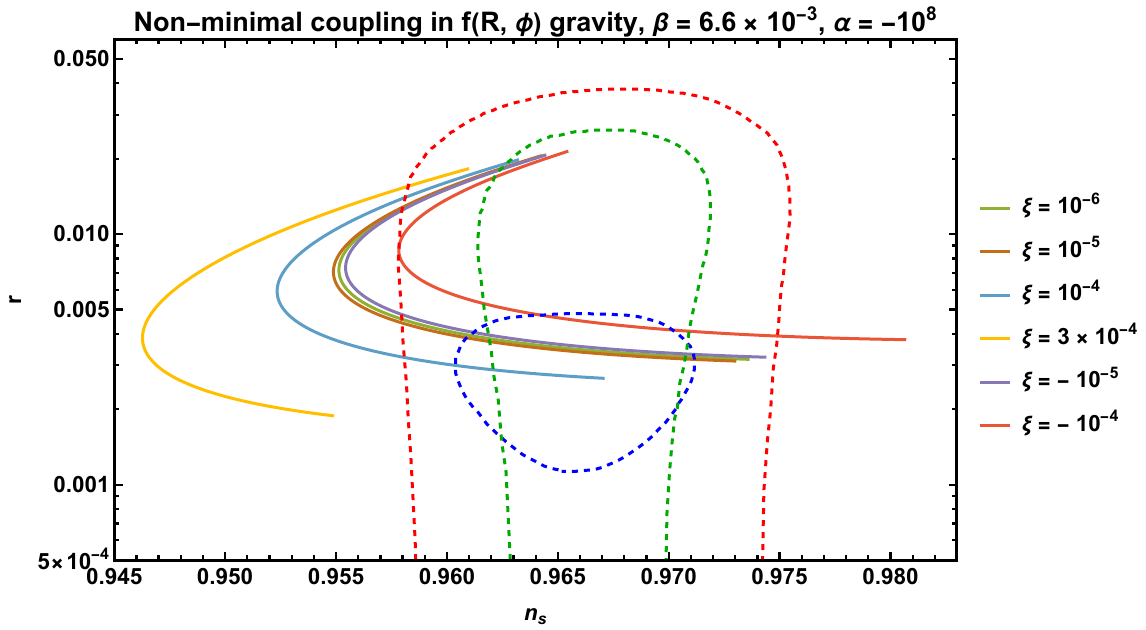}
    \includegraphics[width=0.48\textwidth]{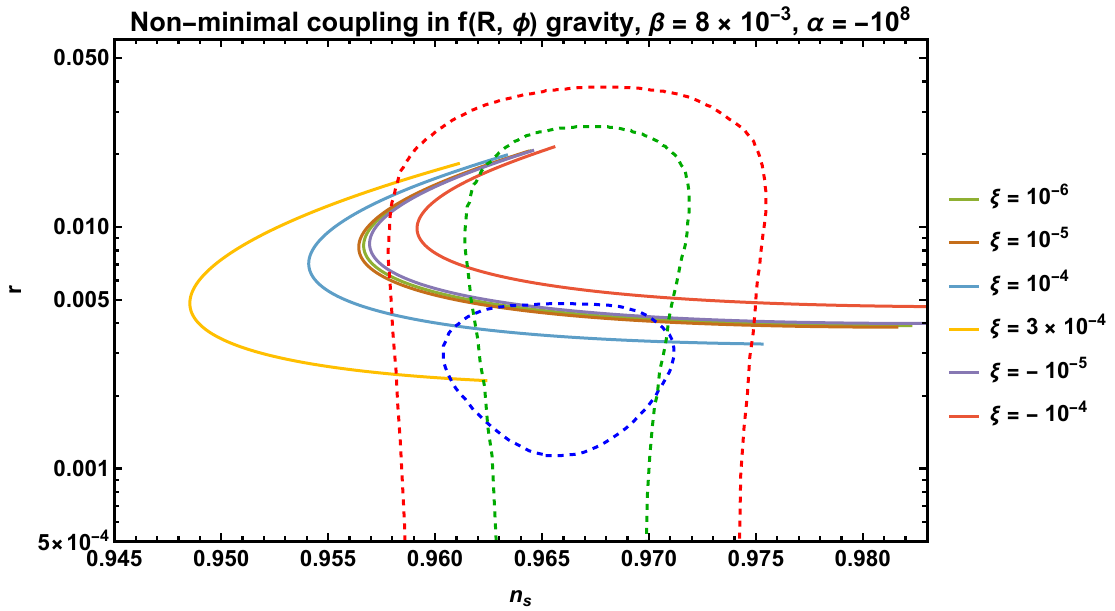}
    
    \
    
    \includegraphics[width=0.88\textwidth]{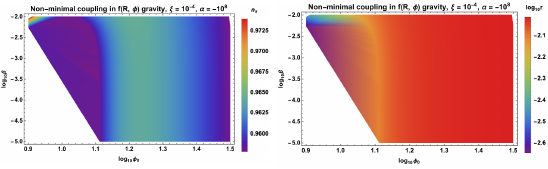}
    \caption{\label{fig:7} Top panel shows the $n_s-r$ predictions for the polynomial inflation with non-minimal coupling in $f(R,\phi)$ gravity for $\beta = 6.6 \times 10^{-3}$ (left pane) and $\beta = 8 \times 10^{-3}$ (right pane) for $\alpha = -10^8$ and selected $\xi$ values. Red (green) dashed contours indicate the recent $95\% \: (68\%)$ CL given by BICEP/Keck \cite{BICEP:2021xfz} and the blue dashed contours correspond to the sensitivity forecast for the future CMB-S4 \cite{Abazajian:2019eic}. Bottom figures show $\beta-\phi_0$ planes for which $n_s$ (left pane), $r$ (right pane) predictions are within the recent $95\% \: (68\%)$ BICEP/Keck CL contours for $\xi=10^{-4}$ and $\alpha=-10^8$. }
\end{figure}
%
%
\subsubsection{Non-minimal coupling} \label{subsubec.4.2.2}  

\begin{figure}[htbp]
\centering
    \includegraphics[width=0.48\textwidth]{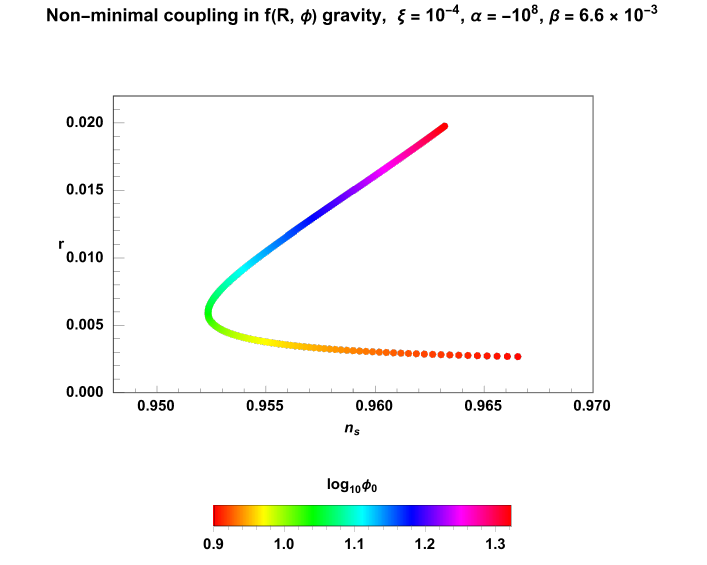}
    \caption{\label{fig:8}  Non-minimal coupling in $f(R,\phi)$ gravity. The relationship between  the $n_s-r$ vs $\phi_ 0$ for $\beta = 6.6 \times 10^{-3}$, $\xi=10^{-4}$ and $\alpha=-10^{8}$.}
\end{figure}

\begin{figure}[tbp]
\centering
    \includegraphics[width=0.53\textwidth]{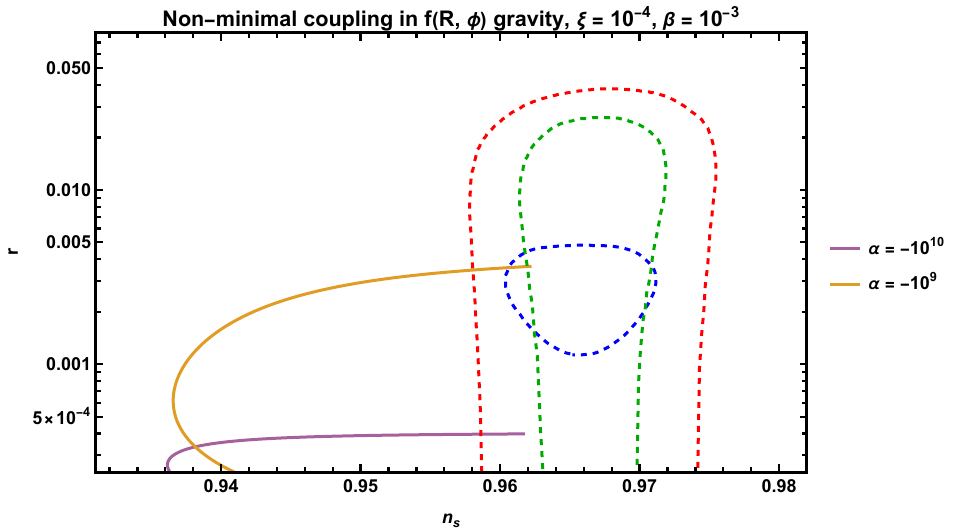}
    \includegraphics[width=0.43\textwidth]{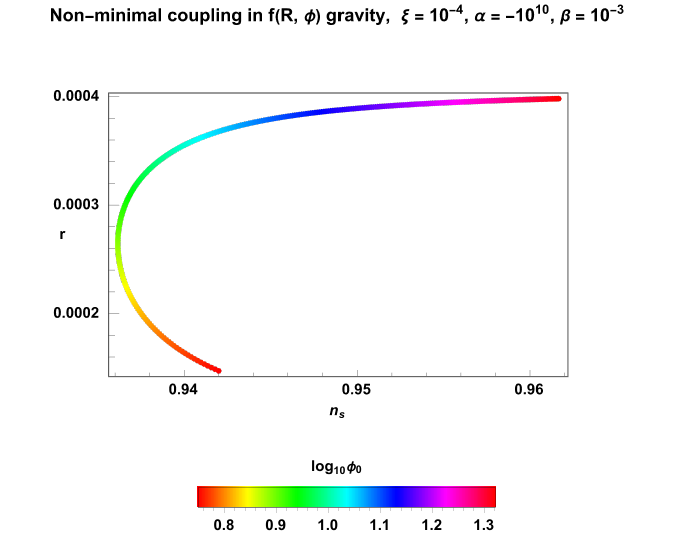} 

    \caption{\label{fig:9} The left pane shows the $n_s-r$ predictions for the polynomial inflation with non-minimal coupling in $f(R,\phi)$ gravity for $\beta = 10^{-3}$, $\xi = 10^{-4}$ and $\alpha=-10^{10}$, $\alpha=-10^{9}$ values. Red (green) dashed contours indicate the recent $95\% \: (68\%)$ CL given by BICEP/Keck \cite{BICEP:2021xfz} and the blue dashed contours correspond to the sensitivity forecast for the future CMB-S4 \cite{Abazajian:2019eic}. The right pane shows $n_s-r$ versus $\phi_ 0$.}
\end{figure}

This section constitutes the final step in the analysis of the PI within the Palatini formulation and it concludes the analysis by incorporating the non-minimal coupling term on top of the Starobinsky term of the previous section. The action of the model is given in (\ref{symmergent-case}) and the potential in (\ref{V_EF}). Figures \ref{fig:7}, \ref{fig:8} and \ref{fig:9} display the results of the numerical analysis for polynomial inflation with non-minimal coupling in $f(R,\phi)$ gravity. Additionally, Table \ref{table:non-minimal_quadratic} lists several model parameters along with the corresponding predictions for the case where $\beta = 6.6 \times 10^{-3}$, $\xi=10^{-4}$ and $\alpha=-10^{8}$.

The top panel in Figure \ref{fig:7} displays $n_s-r$ predictions for a selection of the non-minimal coupling parameter  $\xi$ values. The left and right panes have different $\beta$ values as before. Inspecting both these panes reveals the general effect of  $\xi$ on $n_s  -  r$ which is similar to the non-minimal $R$ gravity case analyzed in Figure \ref{fig:3}. It is observed that an increase in the non-minimal coupling $\xi$ causes a decrease in both $n_s$ and $r$. 

When examining either $\beta = 6.6 \times 10^{-3}$ or $\beta = 8 \times 10^{-3}$ pane, it is evident that the $n_s  -  r$ curves converge to the respective ones with $\alpha=-10^{8}$ drawn in the top panel of Figure \ref{fig:5} as the limit $\xi \rightarrow 0$ is approached.

Our analysis indicates that for both $\beta = 6.6 \times 10^{-3}$ and $\beta = 8 \times 10^{-3}$, the inflationary predictions for the values of \(\xi = 10^{-6}, 10^{-5}, -10^{-5}\) overlap significantly. Furthermore, the expected values of the spectral index $n_s$ and the tensor-to-scalar ratio $r$ may reside within the observationally relevant region.
For example, with $\beta = 6.6 \times 10^{-3}$, $\xi = 10^{-5}$, and $\alpha = -10^8$, we find $N_* \sim 59.36$, $n_s \sim 0.973$, and $r \sim 3.11 \times 10^{-3}$ for $\phi_0 \sim 7.94$.
For larger values of $\phi_0$, approximately 18.77, we find that $N_* \approx 60$, $n_s \approx 0.9626$, and $r \approx 1.84 \times 10^{-2}$ for $\beta = 6.6 \times 10^{-3}$, $\xi = 10^{-5}$, and $\alpha = -10^8$.

For the case of $\xi = 3\times 10^{-4}$, the predictions are ruled out for $\beta = 6.6 \times 10^{-3}$ when $\phi_0$  is in the range $\phi_0 \sim (7.94, 18)$. However, the inflationary predictions can enter the $2\sigma$ CL as $\phi_0$ increases to the range of $\phi_0 \sim (18, 21)$. In contrast, for $\beta = 8\times 10^{-3}$, the inflationary parameters can fall within the $2\sigma$ CL for both smaller and larger $\phi_0$ values with $\xi = 3\times 10^{-4}$. For example, with $\phi_0 \sim 7.943$ and $\phi_* \sim 7.407$, the $n_s-r$ predictions can be consistent with the $2\sigma$ CL, yielding $n_s \sim 0.9624$, $r\sim 2.3\times 10^{-3}$, and $N_* \sim 59.28$.

Furthermore, for the value of $\xi = 10^{-4}$, our findings indicate that the predictions align well with both the recent BICEP/Keck data and the anticipated sensitivity of CMB-S4 for lower $\phi_0$ values. This holds true for both cases of $\beta = 6.6 \times 10^{-3}$ and $\beta = 8 \times 10^{-3}$.

Similar to the minimal coupling discussed in the previous section, in the bottom panel of Figure \ref{fig:7} regions of $\beta-\phi_0$ plane where the predictions for $n_s$ and $r$ fall outside the observational range are noticed. However, here in the non-minimal case, these regions occupy a smaller area. Specifically, these areas extend up to $\phi_0 \sim 13.2$. For $\phi_0 \lesssim 13.2$, as $\phi_0$ decreases, the $\beta$ values should be increased to satisfy the observational bounds. An illustration of this, concerning the $\beta-\phi_0$ plane, occurs at $\phi_0 \approx 10$ and for $\beta \approx 5 \times 10^{-4}$. In this scenario, we find $n_s \approx 0.96$ and $r \approx 7 \times 10^{-3}$. 

 Additionally, Table \ref{table:non-minimal_quadratic} shows that for the values $\beta = 6.6 \times 10^{-3}$, $\xi=10^{-4}$, $\alpha=-10^{8}$, and $\phi_0 \sim 18.33$, we find $N_* \sim 60$, $n_s \sim 0.961$, and $r \sim 1.72 \times 10^{-2}$. 
 
 Figure \ref{fig:8} gives the $n_s-r$ predictions versus $\phi_0$ relationship and the same trend witnessed in the previous sections is also observed here. As the value of $\phi_0$ increases, the tensor-to-scalar ratio tends to increase as well. However, the spectral index decreases up to a certain point and then starts to increase again. This behavior is commonly observed in all the cases studied.

\begin{table}[!htbp]
\centering
\begin{tabular}{|c|c|c|c|c|c|c|}
\hline
$\phi_0$ & $\phi_{*}$ & $d/10^{-14}$ & $N_{*}$ & $n_s$ & $r/ 10^{-3}$ & $\alpha_s / 10^{-4}$ \\
\hline
7.94  & 7.4675  & 6.47 & 59.3 & 0.9670 & 2.66 & -15.05 \\
8.64  & 7.9050  & 5.52 & 59.4 & 0.9590 & 3.12 & -12.01 \\
9.28  & 8.2928  & 5.08 & 59.5 & 0.9551 & 3.74 & -10.20 \\
10.59 & 9.0185  & 4.64 & 59.6 & 0.9524 & 5.44 & -7.91 \\
11.39 & 9.4143  & 4.47 & 59.7 & 0.9525 & 6.67 & -7.02 \\
12.84 & 10.0605 & 4.17 & 59.8 & 0.9539 & 9.13 & -5.95 \\
13.36 & 10.2698 & 4.06 & 59.8 & 0.9546 & 10.03 & -5.69 \\
14.71 & 10.7548 & 3.76 & 59.9 & 0.9566 & 12.27 & -5.21 \\
15.43 & 10.9868 & 3.58 & 59.9 & 0.9576 & 13.39 & -5.05 \\
16.32 & 11.2477 & 3.37 & 59.9 & 0.9587 & 14.69 & -4.91 \\
17.54 & 11.5665 & 3.09 & 60.0 & 0.9602 & 16.28 & -4.80 \\
18.33 & 11.7518 & 2.91 & 60.0 & 0.9610 & 17.20 & -4.76 \\
19.46 & 11.9928 & 2.66 & 60.0 & 0.9620 & 18.39 & -4.74 \\
20.91 & 12.2638 & 2.38 & 60.0 & 0.9631 & 19.69 & -4.74 \\
\hline
\end{tabular}
\caption{Sample set of model parameters and corresponding predictions for the non-minimal coupling in $f(R,\phi)$ gravity with $\beta = 6.6 \times 10^{-3}$, $\xi=10^{-4}$ and $\alpha=-10^{8}$.}
\label{table:non-minimal_quadratic}
\end{table}

The left pane of Figure \ref{fig:9} illustrates how the predictions for $n_s-r$ change in response to variations in the Starobinsky coupling, $\alpha$. The trend we identified in the minimal coupling in Figure \ref{fig:5} is evident here as well. It is observed that increasing the absolute value of $\alpha$ allows for the suppression of the tensor-to-scalar ratio $r$ without limit. Notably, this suppression occurs with minimal impact on the spectrum of scalar perturbations. It is important to note that, although the parameter space is limited than before, it is still possible to obtain reasonable predictions that fall within the observational region for values of $\alpha = -10^{10}$ and $\alpha = -10^{9}$, while using \(\xi = 10^{-4}\) and \(\beta = 10^{-3}\).

The right pane of Figure \ref{fig:9}, on the other hand, illustrates how the $n_s-r$ predictions change depending on the $\phi_0$ values. It can be seen that for smaller $\phi_0$ values, the predictions are ruled out. On the contrary, for larger $\phi_0$ values, the parameters can be inside the confidence levels. For instance, with $\xi = 10^{-4}$, $\beta = 10^{-3}$ and $\alpha=-10^{10}$; for larger $\phi_0 \sim 17.8$, the predictions turn out to be $n_s \sim 0.962$ and $r\sim 4\times 10^{-4}$, however, for smaller $\phi_0 \sim 6.31$, they become $n_s \sim 0.942$ and $r\sim 1.4 \times 10^{-3}$.   

This marks the end of the analysis of PI in Palatini formalism. In the next section, we summarize the results and discuss the implications.


\section{Summary and Conclusions} \label{conc}

In the present paper, the Palatini formalism is utilized to explore the behavior of large field inflation within the framework of $f(R,\phi)$ gravity for a renormalizable polynomial inflaton potential assuming instant reheating. The analysis involves studying the dynamics of the inflaton under different couplings, including minimal and non-minimal coupling in $R$ gravity, as well as  $f(R,\phi)$ gravity. A thorough parameter space scan is conducted to identify inflationary predictions ($n_s$ and $r$) that align with the Planck and BICEP/Keck 2018 results, as well as the projected sensitivity of the future CMB-S4. For all cases, the compliant regions in the $\phi_0-\beta$ plane are illustrated, where $\phi_0$ and $\beta$ represent two crucial parameters of the polynomial inflation model governing the potential's saddle point and flatness in that region, respectively.

The systematic analysis begins with the simplest case, which is the metric-Palatini $R$ gravity. As expected, the results are parallel to those presented in \cite{Drees:2022aea}, which explores large-field polynomial inflation within the metric formulation. The analysis of this minimal case reveals that an increase in the saddle point value, $\phi_0$, leads to a corresponding rise in the tensor-to-scalar ratio $r$. Similarly, an increase in the PI flatness parameter $\beta$ is found to elevate the spectral index. These findings align with the results reported in the literature before.

\begin{figure}[tbp]
\centering
    \includegraphics[width=0.86\textwidth]{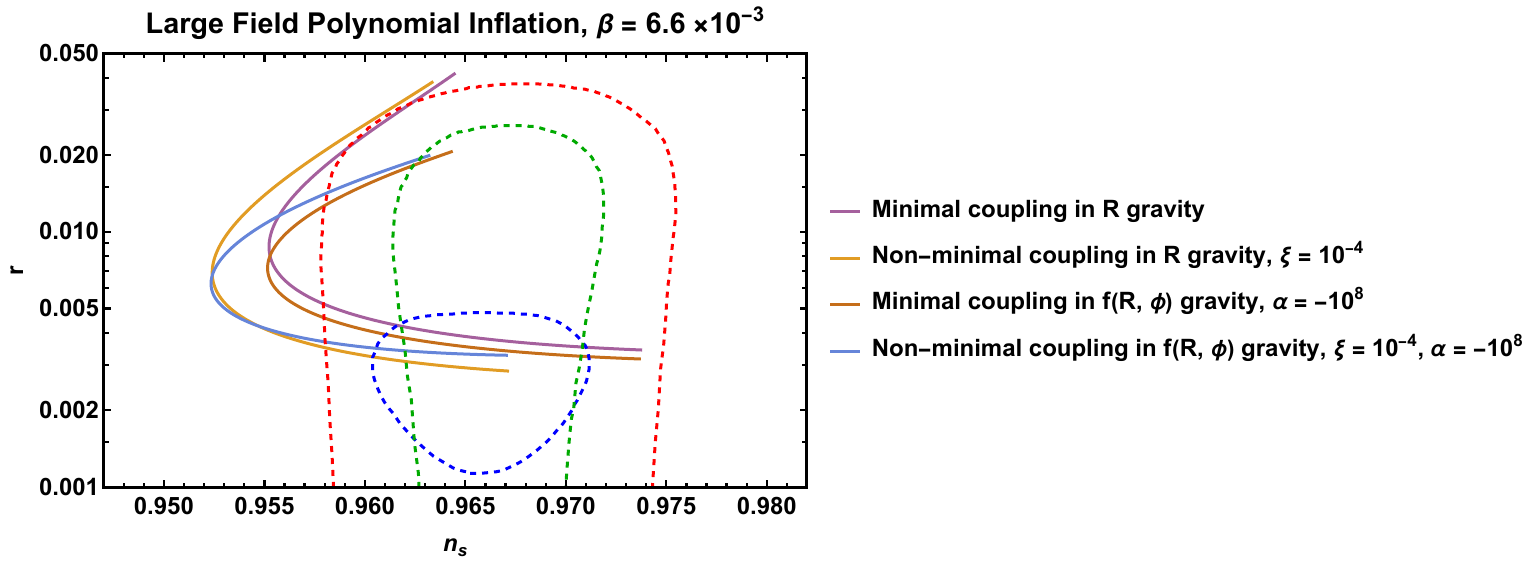}
    \caption{\label{fig:10} Minimal and non-minimal couplings of inflaton in metric-Palatini $R$ and $f(R,\phi)$ gravity.  
    Comparative plot of the $n_s-r$ predictions.
    }
\end{figure}

The second phase of the analysis includes the non-minimal coupling in the metric-Palatini $R$ gravity. The general effect of the non-minimal coupling is that it lowers both $n_s$ and $r$ compared to the minimal coupling in $R$ gravity. This situation is observed in Figure \ref{fig:10} which compares the effects of different couplings on the $n_s-r$.

The third phase involves minimally coupled inflaton in metric-Palatini $f(R,\phi)$ gravity. This phase extends the minimally coupled $R$ gravity by incorporating the $\alpha R^2$ Starobinsky term. The impact of the Starobinsky term in Palatini formalism has been previously explored in the literature, notably in \cite{Enckell:2018hmo}.
Our analysis reveals that the Starobinsky coupling $\alpha$ has the potential to suppress the tensor-to-scalar ratio without limit while leaving other parameters mostly unchanged. This finding is consistent with the results reported in \cite{Enckell:2018hmo} and is illustrated in Figure \ref{fig:10}. The brown curve representing minimal coupling in $f(R,\phi)$ gravity shows lower tensor-to-scalar ratios compared to the purple curve of minimal coupling in $R$ gravity. However, the spectral index values for both remain essentially the same. As reported in Sec. \ref{subsubsec.4.2.1}, increasing the absolute value of $\alpha$ further reduces the tensor-to-scalar ratio. 

The final phase of the analysis examines the non-minimal coupling in $f(R,\phi)$ gravity. It is observed that the effect of this non-minimal coupling is similar to that in $R$ gravity. Specifically, the non-minimal coupling reduces both the spectral index $n_s$ and the tensor-to-scalar ratio $r$ compared to the minimal coupling in $f(R,\phi)$ gravity.

The extensive analysis summarized above reveals that substantial regions within the parameter space consistently align with the observational data. As a result, polynomial inflation in the Palatini formalism emerges as a viable model of inflation and warrants further research in this field. One potential direction for future research involves investigating the (p)reheating process and the formation of primordial black holes within this framework. Another promising area to explore—one that we plan to pursue—is the examination of a Beyond the Standard Model (BSM) scenario that introduces an inflaton, specifically a singlet scalar, along with a higher-spin fermion that could act as a candidate for dark matter. Furthermore, it may be worthwhile to investigate a different scenario where this higher-spin fermion generates CP asymmetry, which could, in turn, pave the way for leptogenesis.


\acknowledgments

We dedicate this work to the memory of Durmu\c{s} Demir, whose unwavering passion, exceptional contributions to physics, and enduring spirit will continue to inspire us.






\end{document}